\documentclass[traditabstract]{aa}
\usepackage{graphicx}
\usepackage{txfonts}
\usepackage{psfrag}
\usepackage{color}
\usepackage{natbib}

\begin{document}
\title{LIME - a flexible, non-LTE line excitation and radiation transfer method for millimeter and far-infrared wavelengths}
\titlerunning{LIME - a flexible, non-LTE line excitation and radiation transfer method}

\author{C.~Brinch \and M.~R.~Hogerheijde}
\date{}
\institute{
  Leiden Observatory, Leiden University, 
  P.O.~Box 9513, 2300 RA Leiden, The Netherlands\\ 
  \email{brinch@strw.leidenuniv.nl}
}
\date{Received  / Accepted }
\abstract{We present a new code for solving the molecular and atomic excitation and radiation transfer problem in a molecular gas and predicting emergent spectra. This code works in arbitrary three dimensional geometry using unstructured Delaunay latices for the transport of photons. Various physical models can be used as input, ranging from analytical descriptions over tabulated models to SPH simulations. To generate the Delaunay grid we sample the input model randomly, but weigh the sample probability with the molecular density and other parameters, and thereby we obtain an average grid point separation that scales with the local opacity. Our code does photon very efficiently so that the slow convergence of opaque models becomes traceable. When convergence between the level populations, the radiation field, and the point separation has been obtained, the grid is ray-traced to produced images that can readily be compared to observations. Because of the high dynamic range in scales that can be resolved using this type of grid, our code is particularly well suited for modeling of ALMA data. Our code can furthermore deal with overlapping lines of multiple molecular and atomic species.}
\maketitle

\section{Introduction}
The ability to predict the radiation signature associated with a physical model is crucial when interpreting astronomical observations. In the general case, this can be achieved by solving the equation of radiation transport and, in the special case of non-local thermodynamical equilibrium (non-LTE) radiation, the equilibrium quantum state distribution. This is a strongly coupled problem that can only be solved by numerical analysis except for the limiting cases where either the material is in LTE or in the optically thin regime where there is little interaction between matter and radiation. The computational demand increases rapidly with complexity of the model if such approximations cannot be applied. Due to the low density conditions often found in inter- and circumstellar material, LTE is oftentimes not a valid approximation and therefore, if the medium turns optically thick, numerical methods must be used to calculate the molecular excitation in such environments.

Ever since the first molecules where discovered in space~\citep[CH$^+$ and OH:][]{Douglas1941,weinreb1963}, line emission has been an important astrophysical tracer of the physics and chemistry. The chemistry can be traced from the abundance of various molecular species and the physical environment, such as temperature and density, from the excitation of the lines and kinematical information is embedded in the line profiles. In star forming regions, molecules are also tracers of shocks and can be used to investigate the external radiation environments~\citep{vandishoeck2009}. As an example, the inside-out collapse model of low-mass star formation~\citep{shu1977}, which is often used by many authors, has been confirmed for a number of objects by measurements of various molecular lines~\citep[e.g.,][]{walker1986, zhou1993}. However, in order to derive reliable molecular abundances, it is important to understand the spatial distribution of the gas, and while the gas distribution in the plane of the sky is directly measurable, given sufficient instrument resolution, the distribution along the line of sight is not easily measured. In fact, deriving the three dimensional distribution based on a two dimensional image forms an inverse problem similar to the problems encountered in medical tomography and seismology. In the astrophysical case, we deal with such problems by constructing a source model based on our theoretical understanding of the object in question, and then calculate the line signature of this model which is then compared to the observations and, based on the result, the model is improved to give a better fit. 

A number of numerical codes and algorithms that focus on predicting molecular emission in the far-infrared and millimeter wavelength regimes have been made available to the community in the last decades~\citep[e.g.,][]{bernes1979,juvela1997,hogerheijde2000a,schoier2000,pavlyuchenkov2004}. As computers have become faster, codes have evolved from working predominantly in a spherical symmetry (1D) to being able to solve problems in a cylindrical symmetry (2+1D). This progress has mainly been driven by the development in telescope facilities, providing observers with ever increasing spatial resolution. This paper describes a new non-LTE radiation transfer code for (sub-)millimeter and far-infrared continuum and spectral line radiation that works in arbitrary geometries (3D). The code has been designed to be fast, reliable, and easy to use, with particular emphasis on being able to solve models in very high resolution for the Atacama Large Millimeter Array (ALMA) and models of molecules with a complex level configuration (e.g., H$_2$O and CH$_3$OH), relevant for observations with the Herschel Space Observatory. The code is called \emph{LIME} (Line Modeling Engine).

\emph{LIME} derives from the \emph{RATRAN} code \citep{hogerheijde2000a} and, although rewritten from scratch, it shares much of the code base and the solution method. \emph{RATRAN} is a Cartesian grid based Monte Carlo code which includes elements of accelerated Lambda iteration~\citep{rybicki1991} and it exists in both a one and two dimensional version. The main difference between the two codes is the photon propagation method (see Sect.~\ref{ss:grid}) which dramatically improves the calculation time, making 3D models feasible to solve, and allows for a very flexible model input method. Our photon transport method is a line extension of the approach of \citet{ritzerveld2006} for their continuum radiation transfer algorithm \emph{simpleX}. This algorithm was originally developed in to provide a fast solution to the radiation transfer problem in three dimensions for all opacity regimes, in particular for hydrodynamical cosmology simulations, where approximate radiation transfer is often used to speed up temperature calculations. The \emph{simpleX} algorithm has been shown to work well for clumpy and inhomogenous media with a large density contrast. \emph{LIME} has been written mainly for the purpose of solving models of disks and envelopes around young stellar objects (YSOs), but due to the 3D nature of the code, it can, however, also solve models of more complex structures such as outflows, molecular clouds and even clusters of YSOs and their environments. The main limitation to this is the availability of input models. It is to be expected that models provided by (magneto-) hydrodynamical simulations will become more important in the near-future, and it has therefore been important for us to provide an easy way of interfacing the output of a dynamical simulation with this code. SPH simulations are in particular well suited as input for \emph{LIME}, since the physical quantities is already described by particles in very much the same manner as \emph{LIME} uses it. 

A number of new features have been added to \emph{LIME}. As it is expected that ALMA will detect potentially hundreds of lines per setting and many of these lines may be blended, we have made it possible to solve the radiation transfer for an unlimited set of molecular species simultaneously and dealing with the cross-excitation correctly. This feature obviously comes at a cost of increased calculation time, of which the severeness depends on the number of coupled lines. Another new feature is the extended output capabilities, where the user can ask for spatial intensity and opacity distributions as well as the grid information be written out for visualization purposes. 

The outline of this paper is as follows: Section~\ref{physics} describes the physical problem of radiation transport in a molecular medium, Sect.~\ref{method} provides an extensive description of the \emph{LIME} code, Sect.~\ref{examples} brings a few examples and validation of the \emph{LIME} results, and finally, conclusions are given in Sect.~\ref{conclusion}.

\section{The physical problem}\label{physics}
The spectral intensity $I_\nu$ of radiation propagating through a medium with emission and absorption coefficients $j_\nu$ and $\alpha_\nu$ is described by the equation
\begin{eqnarray}\label{radtran}
\frac{dI_\nu}{ds} = j_\nu - \alpha_\nu I_\nu,
\end{eqnarray}
where we neglect the scattering term which is of little relevance in the sub-millimeter regime. The coefficients $j_\nu$ and $\alpha _\nu$ are related to the Einstein coefficients $A_{12}, B_{12}$, and $B_{21}$, for transitions between any two adjacent levels with an energy of $h\nu =E_1-E_2$, by the equations 
\begin{eqnarray}
j_{\nu;gas} &=& \frac{h \nu}{4\pi} n_2 A_{12} \phi(\nu) \\
\alpha _{\nu;gas} &=& \frac{h \nu}{4\pi} \left (n_1 B_{12}- n_2 B_{21} \right ) \phi(\nu),\label{alpha_gas}
\end{eqnarray}
for the line radiation and
\begin{eqnarray}
j_{\nu;dust} &=& - \alpha _{\nu;dust} B_\nu(T_{dust}) \\
\alpha _{\nu;dust} &=& \kappa _\nu \rho _{dust},
\end{eqnarray}
for the thermal dust emission. $\phi$ is a function of frequency which contain the contribution of several line broadening mechanisms and $h$ is Plancks constant. The dominating mechanism here is the doppler broadening due to local turbulence. $B_\nu$ is the Planck function for a given dust temperature and $\kappa$ and $\rho$ are the dust opacity and dust mass density respectively. The dust opacities depend on the type of dust and various tabulated descriptions can be found throughout the literature~\citep[e.g.,][]{ossenkopf1994} The contribution from the gas and the dust components add up to form the total emission and absorption coefficients. 

At any given position, the local mean radiation field can be obtained by solving Eq.~\ref{radtran} and integrating the intensity over all solid angles
\begin{eqnarray}
J_\nu = \frac{1}{4\pi} \int I_\nu d\Omega.\label{jbar}
\end{eqnarray}
The gas can be excited either through absorption of a photon or through collisions with other gas molecules. Collision rates between two energy levels $i$ and $l$ depend on the local kinetic gas temperature
\begin{eqnarray}\label{collrates}
C_{il} = \frac{g_l}{g_i}c_{il} \exp{\left (-\frac{hc}{k_B T_{gas}}(E_l-E_i)\right )},
\end{eqnarray}
where $g_i$ is the statistical weight of the $i$'th level and $c_{il}$ is the molecule dependent rate coefficients, which, in general, are functions of temperature as well. $k_B$ is Boltzmanns constant. The molecular excitation thus depends on $J_{il}$ (where $il$ replaces $\nu$ to denote the frequency associated with transition between level $i$ and $l$) and $C _{il}$ and the fractional population of the $i$'th level $n_i$ is given by
\begin{eqnarray}
n_i = \frac{\Sigma_{l>i}n_k A_{li}+\Sigma_{l \neq i}n_l(B_{li}J_{li} + C_{li})}{\Sigma_{l<i} A_{il} + \Sigma_{l\neq i} (B_{il}J_{il}+C_{il})},\label{matrix}
\end{eqnarray}
assuming statistical equilibrium. Because $J_{il}$ implicitly depend on $n_i$ through $j_\nu$ and $\alpha_\nu$, Eq.~\ref{jbar} and~\ref{matrix} forms a recursive problem and must be solved iteratively. The collisional rates $c_{il}$ in Eq.~\ref{collrates} are not provided as part of the \emph{LIME} package and must be obtained separately. \emph{LIME}, however, can read the collisional data files provided by the LAMDA database\footnote{http://www.strw.leidenuniv.nl/$\sim$moldata}.

The whole problem is further complicated if a systematic velocity field is present. In this case photons will be Doppler shifted and thereby contribute to different transitions if the transition energies are spaced closely enough. While this is not a problem for molecules with sufficiently widely spaced transitions, such as CO, more complex molecules such as CH$_3$OH have a much richer level sub-structure where photons can easily contribute to several transitions given a systematic velocity field or even by random velocities in the gas. Indeed, a photon originating from a transition in one molecule may get Doppler shifted and excite a transition in another molecule, and therefore we track photons by their frequency, not relative velocity, for each transition in question, summing up the contribution to $J_\nu$ from all other sufficiently close transition. 

\section{Solution method}\label{method}
\subsection{Computational grid}\label{ss:grid}
In  radiation transfer codes, the source model is typically laid out onto either a grid of rectangular cells or an $(r,\theta)$--grid. Model properties such as density, temperature, molecular abundance, as well as the radiation field $J_\nu$ and the level populations $n_i$ are taken to be constant over each cell. Photon packages are traced in random directions from random points of absorption within a cell and Eq.~\ref{radtran} is integrated along the paths. Summing over all photons, the mean radiation field $J_\nu$ is obtained for the cell.  

In our implementation, the source model is not mapped onto a regular grid of cells. Instead we use a random set of points which represents the local environment (density, temperature, populations, etc.). The points are distributed in three-dimensional space, and thus we are able to use three-dimensional source models whereas in, for instance \emph{RATRAN}, even though photons propagate in three-dimensional space, the source model needs to be rotational symmetric around the second axis and mirror symmetric around the first axis. Our points are placed randomly throughout the entire computational domain, however with a probability that is weighted by a source model dependent function. This approach is similar to the one described by~\citet{ritzerveld2006}. We choose to use the molecular density profile of the source model as a probability distribution for the grid points, so that we end up with a point distribution that has particularly interesting properties from a radiation transfer point of view, namely, that the average distance from a point to its neighbors becomes inversely proportional to the density and proportional to the local mean free path, since the mean free path $l = (\alpha _\nu \rho)^{-1}$. 

If we at first consider continuum radiation transfer only, where the absorption coefficient is independent of the radiation field, $\alpha _\nu$ is constant and the mean free path depends on density alone. We can thus obtain a grid point distribution where the expectation value of the neighbor point separation equals the local mean free path by adjusting the number of grid points in accordance to the dust opacity $\kappa _\nu$.

\begin{figure}[!t]
  \begin{center}
	\includegraphics[width=8.8cm]{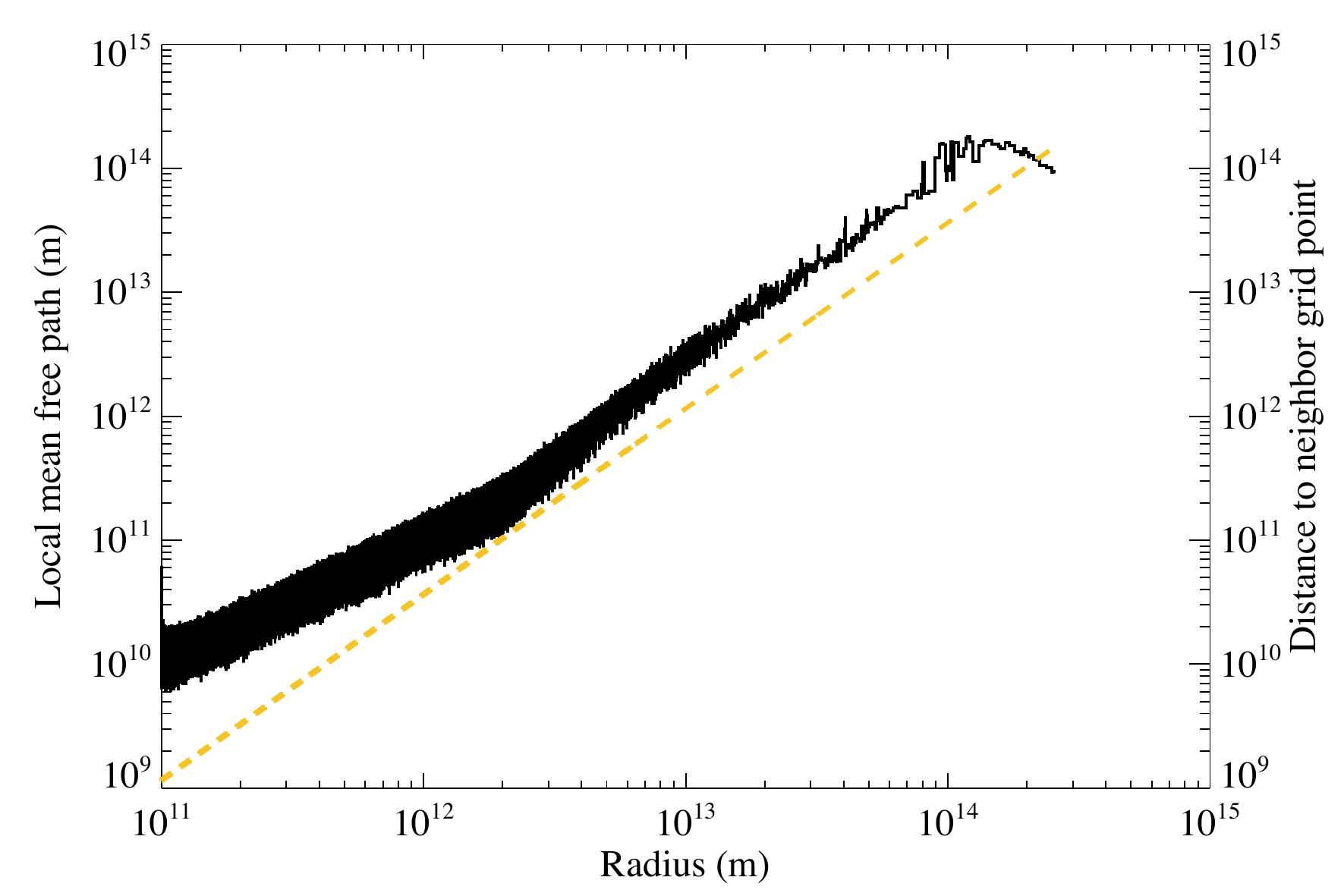}
\end{center}
  \caption{The dashed line shows the local mean free path in the continuum for a model with a density profile $\rho \propto r^{-1.5}$ and a constant absorption coefficient. The full line is the average neighbor distance for a random density weighted point distribution.}\label{mfp}
\end{figure}

Figure \ref{mfp} shows that the average neighbor distance is proportional to the density over two orders of magnitude. The arbitrary offset between the mean free path graph and the point separation graph scales with the total number of grid points. The point separations are seen to deviate from proportionality at radii smaller than $10^{12}$ m ($\approx 7$ AU). This is because we impose the constraint that a certain number of points should be present a the largest scales and thus it would take an unfeasible number of points to follow proportionality with the mean free path down to 0.1 AU. The point distribution in Fig.~\ref{mfp} consists of 40000 points. \emph{LIME} offers several different options for sampling the density, depending on the geometry of the input model, such as uniform sampling in a rectangular box or logarithmic radial sampling in a sphere. It is also possible to use a fixed set of points, for instance the cell centers of a tabulated input model or the particles of an SPH simulation. In this case, the resulting point distribution may not, of course, scale with the mean free path. In other cases it may not be desirable to grid according to the molecular density. If, for instance, the user is interested only in the very high $J$-lines of a molecule, it can be useful to grid according to the temperature distribution, or if the model involves shocks or outflows, the velocity gradients. In that case however, the grid does not describe the opacity well, but rather ensures that certain spatial region are well sampled. It is possible to logically OR point distributions, so that a grid may be based on both the temperature and the density, in which case the opacity is also sampled. In fact, point distributions for each individual opacity source (e.g., gas and dust) should always be OR'ed to form a single grid.

If we now consider line radiation transfer, the local absorption coefficient $\alpha_\nu$ depends on the current level population through Eq.~\ref{alpha_gas}. The local mean free path thus changes with the radiation field and no longer scales simply with the density, but rather with the population dependent opacity. Furthermore, the opacity varies across the line, so photons at different frequencies will have a different mean free path, and therefore the local photon mean free path is not a well defined property. Despite of this, gridding according to density turns out to be the best way to form the grid for line radiation transfer too, simply because it results in a grid that describes the spatial distribution of the molecules very well.

\begin{figure*}
  \begin{center}
	\includegraphics[width=18cm]{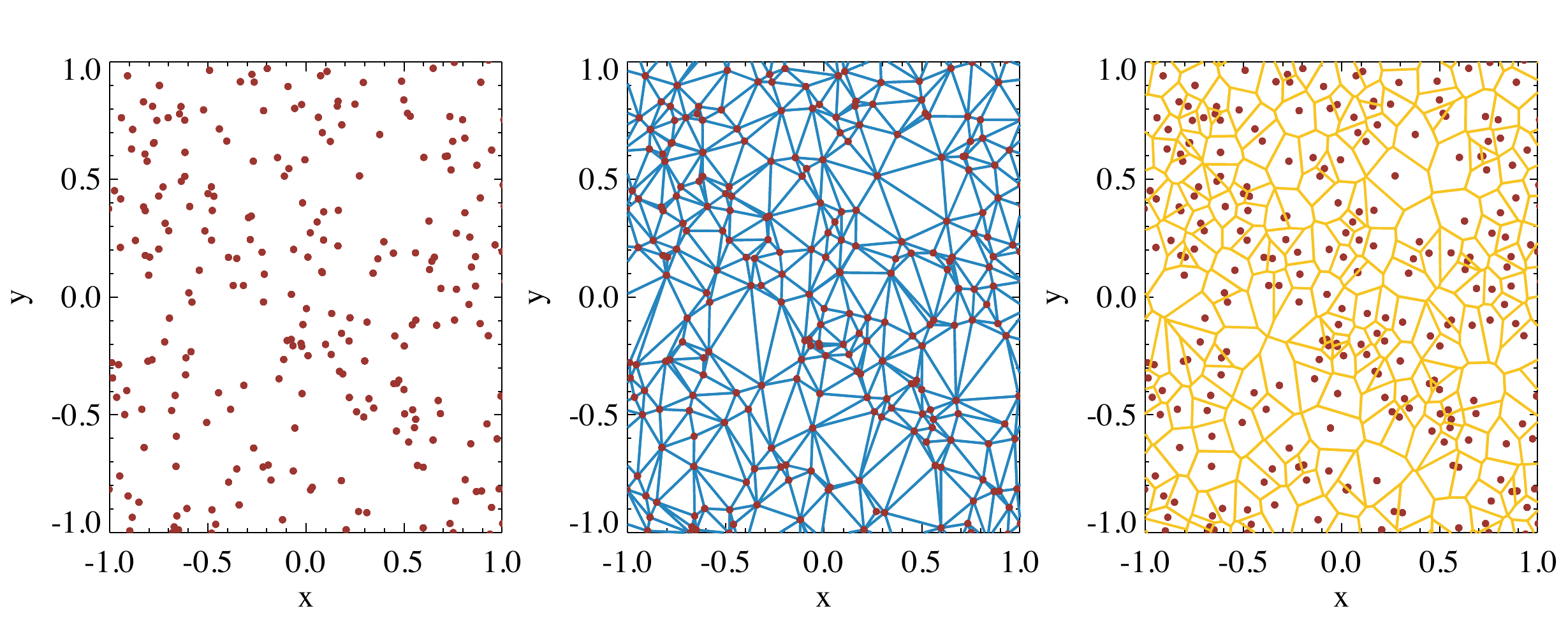}
  \end{center}
  \caption{The leftmost panel shows a random point distribution. In the middle panel, the points have been Delaunay triangulated. The rightmost panel shows the corresponding Voronoi tessellation. }\label{random}
\end{figure*}

When the point distribution has been obtained, the points are connected by Delaunay triangulation, using the public available \emph{QHull} library~\citep{barber1996}. In 2D, the Delaunay triangulation is constructed by connecting any three points that defines an empty circumcircle, meaning that no other points can lie inside the circle defined by a Delaunay triangle. The definition is valid for all higher dimensions and in 3D the Delaunay triangulation forms tetrahedra out of four vertices. Figure~\ref{random} shows a random point distribution (in 2D) in the left panel with its Delaunay triangulation shown in the center panel. Also shown, to the right, in Fig.~\ref{random} is the corresponding Voronoi diagram. The Voronoi diagram is the topological dual of the Delaunay triangulation and one can be constructed from the other. For a discrete set of generating points $P$, a Voronoi cell is defined as the region surrounding the site $s \in P$ which contain points that lie closer to $s$ than to any other generating sites in $P$. The physical properties of the grid points (density, temperature, excitation, etc.) thus represent the entire Voronoi cell associated with that point and therefore the Voronoi cell can be considered similar to the cells in traditional codes. As a matter of fact, if we chose a regularly sampled and unweighted grid point distribution, the Voronoi diagram will describe an ordinary Cartesian mesh. One beneficial property of the Voronoi grid is that it automatically conserves the mass when mapping a density function. This is difficult to do on a Cartesian grid because the mass centroids of the cells are dependent on the cell orientation with respect to the underlying model. Furthermore, because of the random orientation and shape of its cells, Voronoi grids do not suffer from the aliasing effects which are inherent to regular Cartesian grids.

\begin{figure}
  \begin{center}
 	\includegraphics[width=8.8cm]{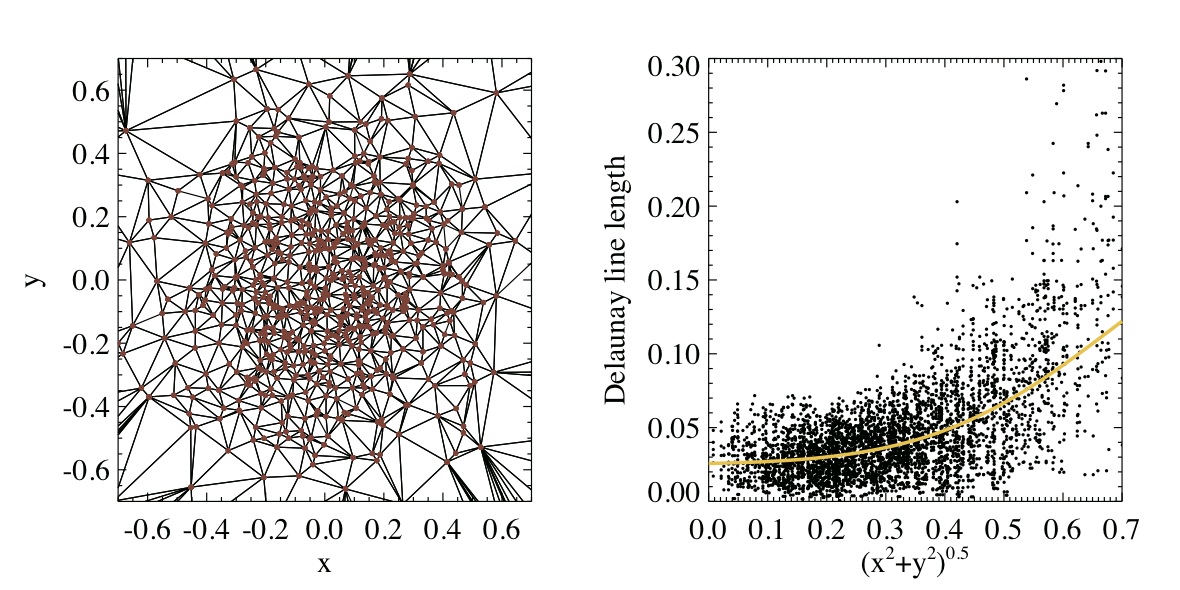} \\
	\includegraphics[width=8.8cm]{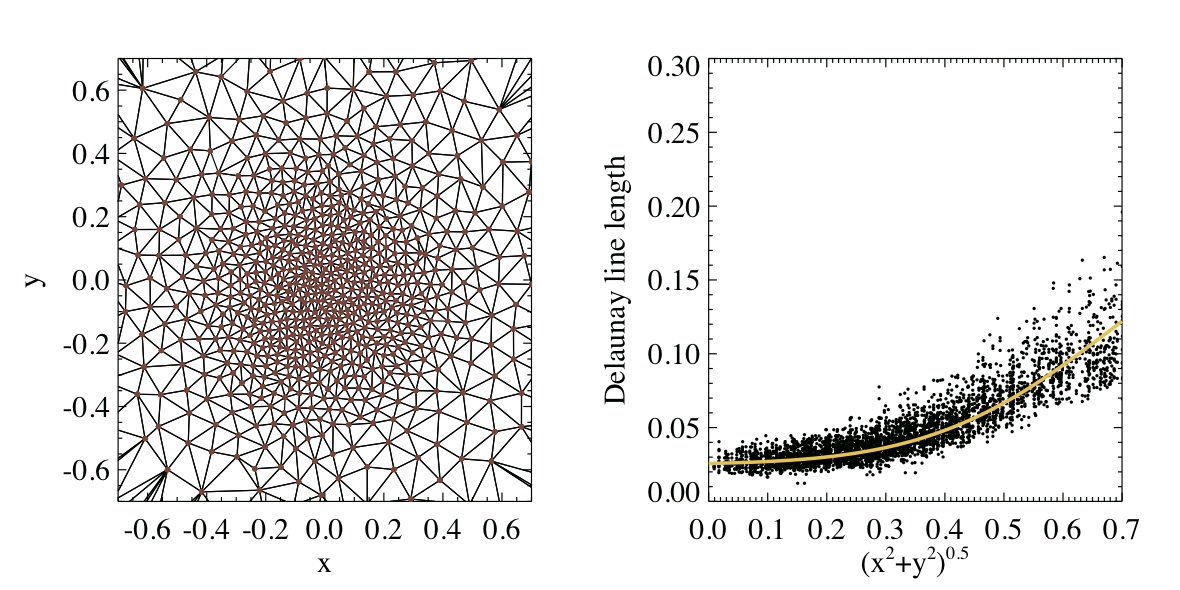}
\end{center}
  \caption{An unsmoothed (top) and smoothed (bottom) Delaunay grid based on a Gaussian density distribution. The right hand column shows the neighbor point separations as function of radius. The yellow lines are aids to the eye. They show a Gaussian distribution that describes the density distribution.}\label{smooth}
\end{figure}

When the grid points have been distributed throughout the model domain, it is inevitable, due to the stochastic nature of the sampling method, that some points end up much closer than the local separation expectation value and some will be much further apart. This results in a Delaunay triangulation that is very irregular with some triangles being very long and narrow. This irregularity can be remedied by applied what is known as Lloyds algorithm~\citep{Lloyd1982,springel2010}, which iteratively moves a grid point slightly toward its Voronoi cells center of mass. In our implementation, each grid point is moved slightly away from its nearest neighbor for a preset number of iterations. The effect is illustrated in Fig.~\ref{smooth} for a random 2D sampling of a Gaussian density profile. The top panels shows the initial unsmoothed distribution while the lower panels showed the triangulated point distribution after the smoothing algorithm has been applied. The plots in the right column show the neighbor distances as a function of radius. In the smoothed grid, the distances are much less scattered and follows the Gaussian profile (shown as the light colored full curve) much more accurately than in the top panel. The smoothing strategy should not be exaggerated, i.e., moving the points too slowly and iterating trough too many stpdf, because the algorithm will then act as an annealing process and it will result in a perfectly regular grid where all variations in the point distribution due to the underlying density field is smeared out. By doing it right, however, a smooth grid can be obtained while the underlying density structure is still preserved in the grid. We have found empirically that by using 25 iterations and moving the closest neighbors about 10\% of the distance away from each other results in a sufficiently smooth grid that preserves the underlying physical structure well.

During gridding of our source model, we also distribute a number of points randomly on the surface of a sphere surrounding our model. These points are also Delaunay triangulated and connected to the model grid points, but they do not represent anything except the surface of our computational domain. Whenever a photon reaches one of these sink points, it is considered to have escaped the model.

\subsection{Photon propagation}
\begin{figure}[!t]
  \begin{center}
	\includegraphics[width=8.8cm]{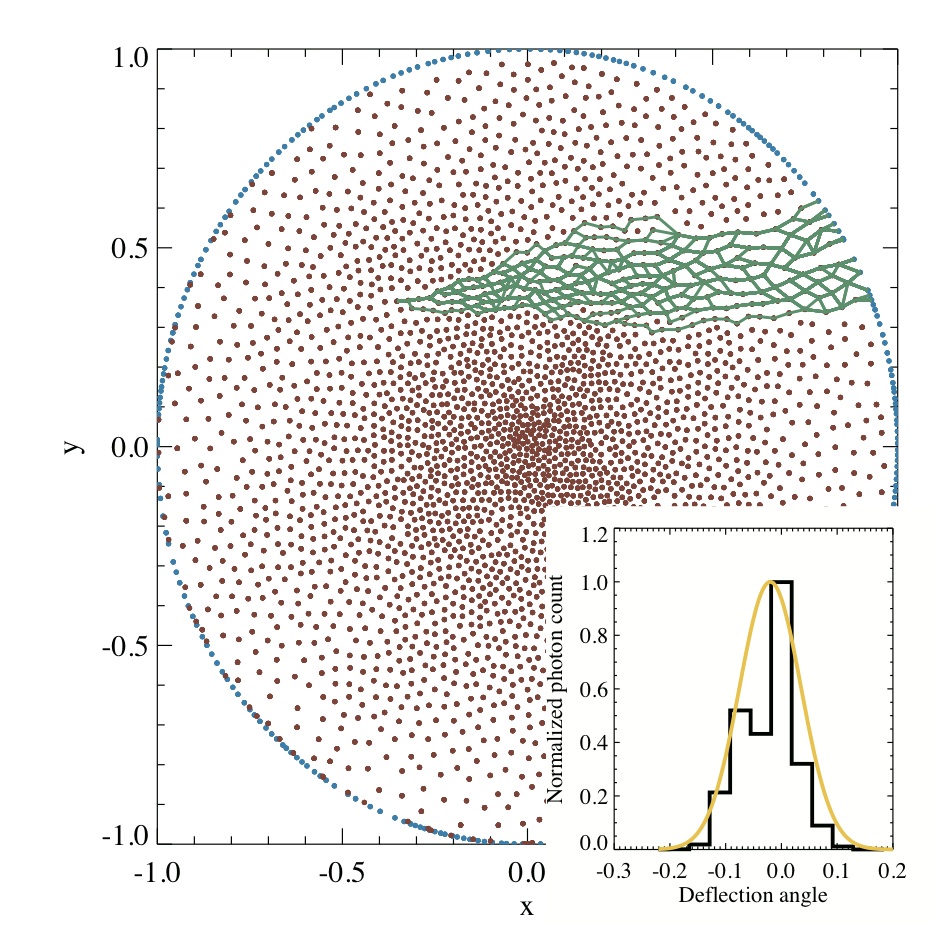}
  \end{center}
  \caption{The propagation of photons from one point in one particular initial direction. As the photons step through the grid they choose the following step by weighing the probability with the inverse angle the direction makes with the initial direction, i.e., a large angle produces a small probability of going in that direction. The inset shows the distribution of photon arriving at the surface. A Gaussian distribution is overplotted for comparison.}\label{photprop}
\end{figure}

The photon transport itself goes along Delaunay lines only, from one point to another, which makes integration of Eq.~\ref{radtran} particularly simple and very fast. In the three-dimensional Delaunay triangulation, the expectation value for the number of lines attached to a grid point is approximately 16 \citep{ritzerveld2006} and the spatial sampling of $J_\nu$ is thus limited to this number of directions. However, we still need to trace a number of photons along each Delaunay line, not only in order to sample the frequency band properly, but also because we cannot conserve momentum stringently with a single photon on this grid. In principle, a photon passing a grid point from a certain direction should continue to travel in the exact same direction. This is in general not possible due to the random orientation of the Delaunay lines, so instead we choose one of the two outgoing Delaunay lines ($\ell1$ and $\ell2$) that make the smallest angle with the original direction of the photon. The outgoing line is picked at random, but weighted by the ratio of the two angles,
\begin{eqnarray}
p(\ell2) = (\angle_1 / \angle_2) p(\ell1), 
\end{eqnarray}
where $\angle_1 < \angle_2$. The same procedure is used at all subsequent grid points (using the original momentum vector to determine the outgoing direction) until the photons escape the model. By sending a number of photons along each initial Delaunay line, we thus probe, not a single line of sight, but rather a cone, while still conserving momentum on average. An example of the photon propagation is shown in Fig.~\ref{photprop} for a single point and a single direction. Because of the relative low number of photons needed to probe the spatial directions, we can allow ourself to increase the number of photons used to sample different frequencies, while we still maintain a low (initial) number of photons per grid point. The inset in Fig.~\ref{photprop} shows the distribution of the location where the photons reach the surface of the grid. This distribution is reasonably well described by a Gaussian distribution around the intersection of the original momentum vector and the surface. The number of initial photons is a user-defined setting, but as a default value, we use five times the number of neighbor points, so that each neighbor is initially probed by five photons. These photons are distributed evenly across a frequency range of $\pm 3 \sigma$ with respect to the line center so that the median photon coincides with the local rest frequency. $\sigma$ is determined by the local turbulent velocity dispersion through the user-defined Doppler b-parameter.
 \begin{figure}
  \begin{center}
	\includegraphics[width=8.8cm]{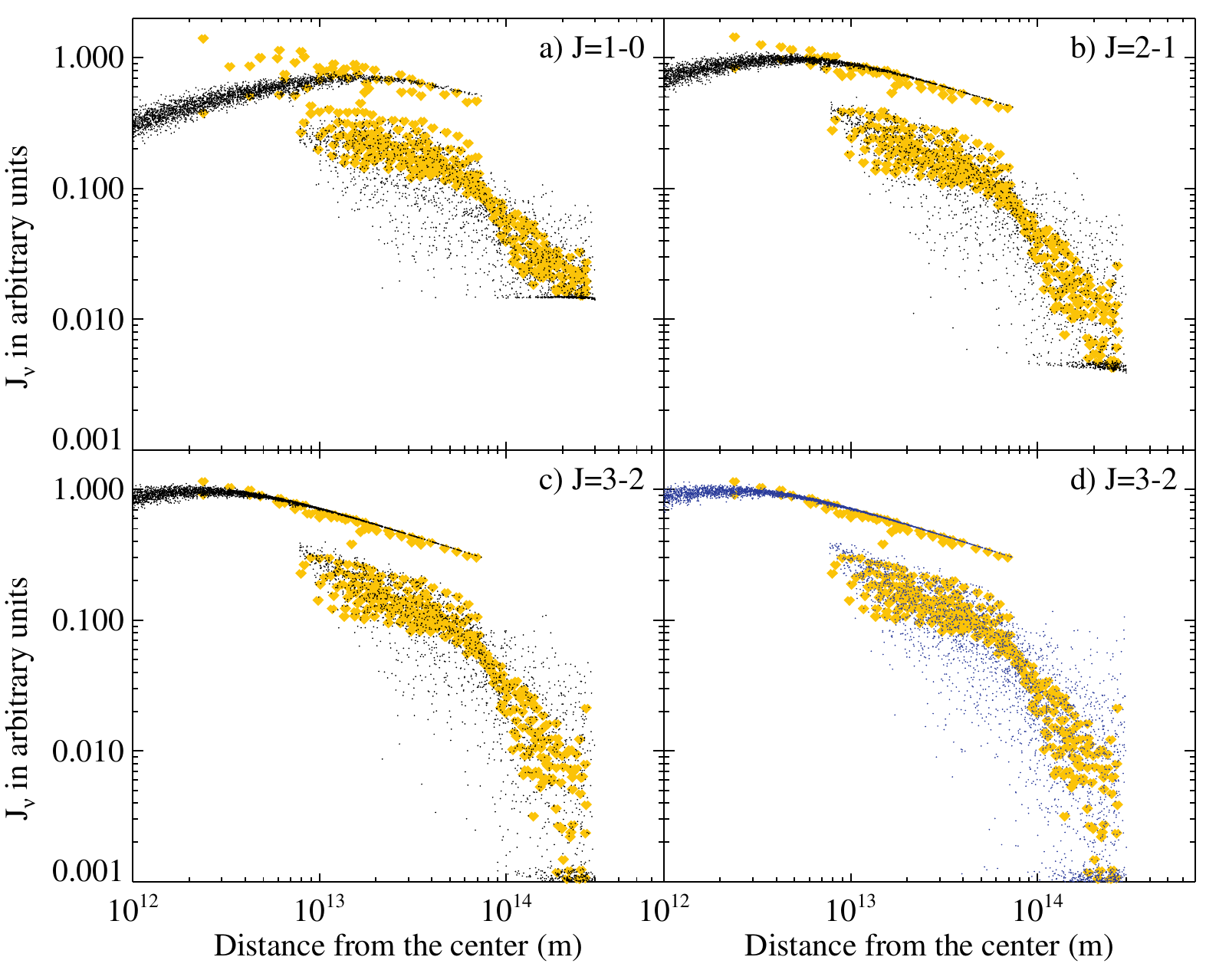}
  \end{center}
  \caption{A comparison of $J_\nu$ between \emph{LIME} and \emph{RATRAN}. The small black dots are the values from the \emph{LIME} code and the yellow dots are the cell values from \emph{RATRAN}. The blue dots in panel d) are also from a \emph{LIME} model, but where all the grid points have been distributed randomly over the model domain (with no density weighting).}\label{flatdisk}
\end{figure}

Any given grid point will see more Delaunay connections coming from high density regions than from low density regions, simply because the grid point density is higher in high density regions. Because of this inhomogeneity in the angular distribution of Delaunay connection, care must be taken when averaging the radiation field using Eq.~\ref{jbar}. In our implementation, this equation reduces to a discrete sum
\begin{eqnarray}
J_\nu = \frac{1}{4\pi} \sum_{\nu} \sum_{i=0}^{N} I_{i,\nu} \omega_i \phi(\nu),
\end{eqnarray}
where $N$ is the number of Delaunay neighbors. $\omega_i$ is a weight that is proportional to the solid angle represented by the $i$'th Delaunay line. This angle corresponds strictly to a surface area on a unit sphere, but we use the area of the Voronoi facet that corresponds to the Delaunay line as a good approximation (within 10\%). 

Figure~\ref{flatdisk} shows a comparison of $J_\nu$ between \emph{LIME} and \emph{RATRAN}. The input model is a thin flat disk with a density profile $\propto r^{-1}$. The radius is 500 AU and the height is 50 AU. The disk is placed in an ambient low density field, n=$10^4$ cm$^{-3}$. For \emph{LIME} the model is sampled by 8000 points, whereas \emph{RATRAN} uses 400 grid cells. The radiation field of first three levels are shown in panel a)-c) in Fig.~\ref{flatdisk}. The \emph{LIME} points are shown as black points and the \emph{RATRAN} points in yellow. The points for each transition makes up two distinct populations, an almost horizontal branch and a scattered population below. The tight horizontal distribution of points are the ones that lie inside the disk. These points are extremely well matched between the two codes. The scattered population of points are the ones that fall outside of the disk radius and these are also well matched. The \emph{LIME} points in the ambient low density region scatters a bit more than the corresponding \emph{RATRAN} points do. This is not a dilution of the radiation field due to poor spatial sampling or erroneous photon propagation on the Delaunay grid, but simply because the \emph{LIME} grid has a much higher resolution than the \emph{RATRAN} grid. The proof of this can be seen in panel d) in Fig.~\ref{flatdisk} where a similar comparison of the $J=$ 3--2 transition between \emph{LIME} and \emph{RATRAN} has been made, but with a \emph{LIME} grid which is not density weighted at all, which means that all spatial regions are equally well sampled. This distribution is indistinguishable from the one in panel c).

In \emph{LIME} we integrate Eq.~\ref{radtran} all the way to the edge of the model at which point we add the external contribution, in most cases the cosmic microwave background. This is know as the method of long characteristics. Another approach exists, known as the method of short characteristics \citep{kunasz1988,auer1994}, where photons are only traced to the neighbor grid cell at which point an interpolated radiation field is added. Short characteristics could be implemented in \emph{LIME} with a possible gain in calculation time. In fact, when we average the radiation field by splitting the photon packages along two outgoing Delaunay lines, it is reminiscent of the short characteristics method.

The level populations $n_l$ in a grid point, as defined by Eq.~\ref{matrix}, represent the molecular excitation in the entire Voronoi cell associated with the point. $J_\nu$ is then the radiation field seen by the grid point. Following the method of~\citet{rybicki1991}, when working with Cartesian grids, the radiation transfer can be formulated iterative as
\begin{eqnarray}
J_\nu^{k+1} = \Lambda [S_\nu (J_\nu^{k})],
\label{lambda}
\end{eqnarray}
where $S_\nu \equiv j_\nu/\alpha_\nu$ is called the source function and $k$ refers to current iteration. $\Lambda$ is a matrix in which each element represents the radiation coupling between each pair of grid points. Thus the diagonal represents the radiative contribution from the cell itself while the off-diagonals are the coupling between the cell and its neighbors. If the entries in the matrix are organized so that cells that are immediate neighbors lie close to the diagonal, it is possible to split $\Lambda$ in the following way
\begin{eqnarray}
J_\nu^{k+1} = \Lambda_{ext} [S_\nu (J_\nu^{k})] + \Lambda_{local} [S_\nu (J_\nu^{k+1})],
\end{eqnarray}
where $\Lambda_{local}$ is a di- or triagonal matrix (depending on the implementation) representing the radiative interaction from the local neighborhood. This will entirely dominate over the external contribution in high opacity problems . Because the $\Lambda_{local}$ matrix is much easier to invert, this approach is much faster than solving Eq.~\ref{lambda} directly and this is in essence what is known as accelerated $\Lambda$-iteration (ALI)~\citep{rybicki1991}.
\begin{figure}[!t]
  \begin{center}
  	\definecolor{mygreen}{rgb}{0,0.69,0}
	\psfrag{Jlocal}[c][c]{\textcolor{mygreen}{$J_{local}$}}
  	\definecolor{myred}{rgb}{0.69,0,0}
	\psfrag{Jext}[c][c]{\textcolor{myred}{$J_{ext}$}}
	\psfrag{celli}[c][c]{cell$_{i}$}
	\psfrag{cellj}[c][c]{cell$_{j}$}
	\psfrag{x}[c][c]{$x$}
	\psfrag{y}[c][c]{$y$}
  	\definecolor{myblue}{rgb}{0,0,1}
	\includegraphics[width=6.8cm]{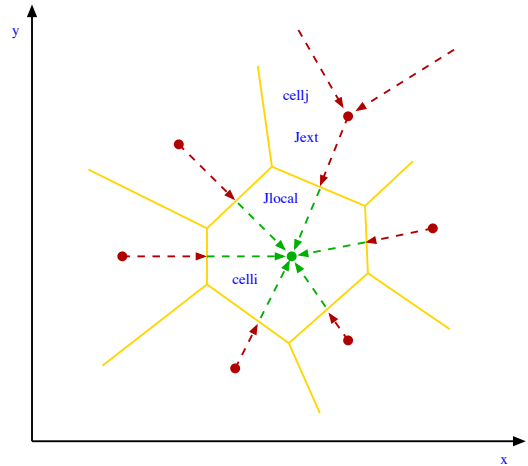}
  \end{center}
  \caption{The radiation field seen by a grid point $i$ is split into a local part $J_{local}$ that originates from within the $i$'th Voronoi cell and an external part $J_{ext}$ that originates from outside the $i$'th Voronoi cell. The size of the Voronoi cells, i.e., the region in which the radiation is considered ``local'', scales with the inverse of the density.}\label{levelpops}
\end{figure}

Because we cannot order the matrix elements of $\Lambda$ in a meaningful way, we do not actually construct the $\Lambda$ operator, but rather split $J_{\nu,total} = J_{\nu,local}+J_{\nu,ext}$, separating the radiation field in the parts that comes from the inside and the outside of the Voronoi cell for which we are solving the populations (Fig.~\ref{levelpops}). This split results in a solution to Eq.~\ref{radtran} that looks like the following,
\begin{eqnarray}
J_\nu = I_\nu e^{-\tau} + S_\nu (1-e^{-\tau}).
\end{eqnarray}
Because $S_\nu$ is a function of $j_\nu$ and $\alpha_\nu$, which are evaluated using the populations in the previous iteration, this method is similar to ALI. $\tau$ is the opacity of the cell itself, so that if $ds$ denotes the distance from the grid point to the Voronoi face $\tau_\nu = \alpha_\nu ds$. In the case where the face of the Voronoi cell is located exactly one mean free path away from the grid point, 
\begin{eqnarray}
\tau_\nu = \alpha_\nu (\alpha_\nu \rho)^{-1} \propto 1/\rho.
\end{eqnarray}
This is exactly the condition we are striving to obtain by gridding according to the density as described in Sect.~\ref{ss:grid}. Of course, the mean free path is frequency dependent, as discussed above, and therefore we can only achieve a condition where $\tau$ scales with inverse of the density. The populations $n_i$ can now be obtained from the current $J_\nu$ by solving the linear set of equations given by Eq.~\ref{matrix},

\subsection{Convergence}
One of the key problems when obtaining an equilibrium solution iteratively is to decide when the solution has converged. There are many ways in which we can check for convergence, but it is necessary to find a method that applies to all models and molecules. It is important that the code does not quit prematurely before the model has actually converged, but on the other hand, it should not continue indefinitely because of minor random fluctuations in a single grid point somewhere. Given a sufficiently large number of grid points, the random nature of the code implies that some points will always deviate from the equilibrium populations. In \emph{LIME} we therefore use a statistical criterion, and let the user decide how many iterations he or she will let the code run. We have found empirically that 15-20 iterations a good value for most models and the default is thus set to 20. This can however easily be adjusted by the user according to preference and the problem at hand.

\begin{figure*}[!t]
\begin{center}
\includegraphics[width=17cm]{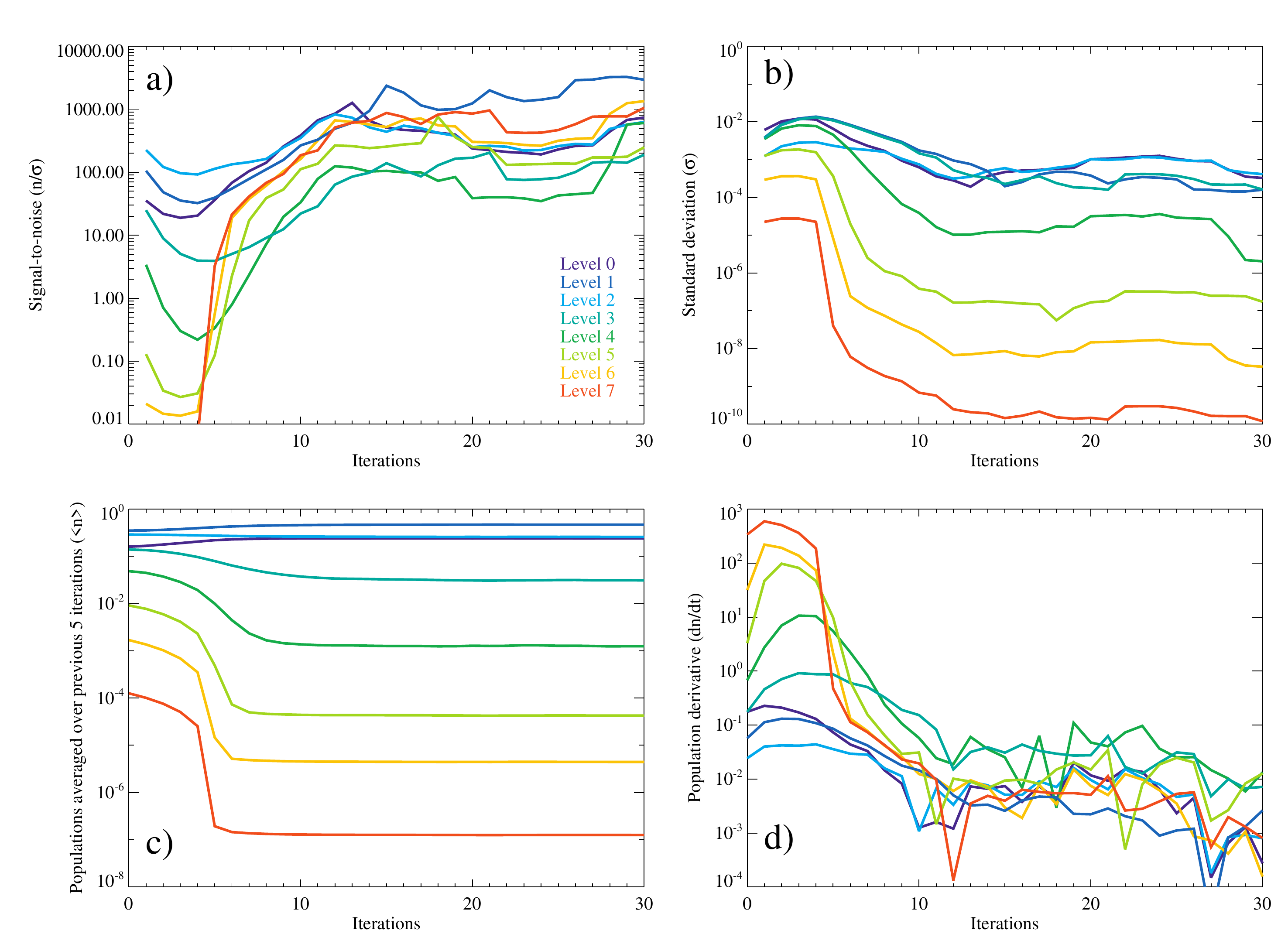} 
\end{center}
\caption{This figure shows the convergence history of a single grid point in a \emph{LIME} run. Different colors refers to the quantum states 0 -- 7. Panel a) shows the signal-to-noise ratio of the current iteration. Panel b) shows the standard deviation, calculated from the previous five iterations. Panel c) shows the average population over the previous five iterations and panel d) shows the derivative of the populations.}\label{statistics}
\end{figure*}

Figure \ref{statistics} shows, as an example, the convergence history of a single grid point in a test run. Panel a) shows the signal-to-noise ratio of the current iteration, where the signal is the current population and the noise is the standard deviation of the previous five iterations (shown in panel b),
\begin{eqnarray}
S/N = \frac{n}{\sqrt{\frac{1}{5} \sum_{j=0}^5 (n_{j}-\bar{n})^2 }},
\end{eqnarray}
for each level. The signal-to-noise ratio is seen to increase and then level out after about 10 iterations. It does however level out at different values for the different levels which makes it difficult to fix a certain value to reach. Also, the signal-to-noise ratio fluctuates a lot (notice the log scale) which again makes it hard to decide whether or not the solution is stable. In panel c) we can see the populations averaged over the previous 5 iterations, and from this plot it is quite obvious that a stable solution (for this grid point) has been reached after the 12'th iteration. The derivative of this curve is shown in panel d). 

\begin{figure*}[!th]
\begin{center}
\includegraphics[width=16.5cm]{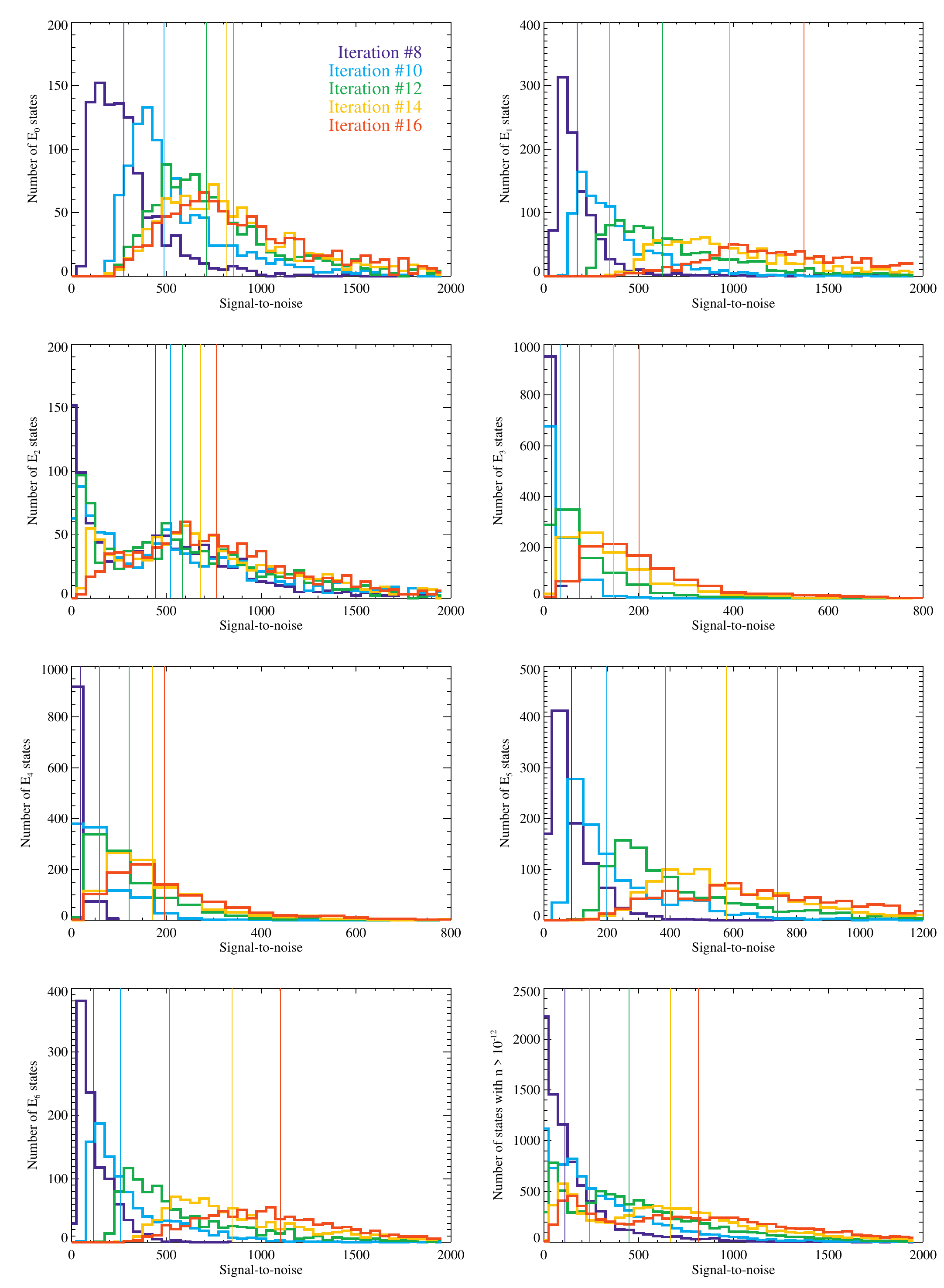} 
\end{center}
\caption{Distributions of signal-to-noise ratios in a test run for energy levels 0 -- 6 as well as for all significantly populated levels. The color coding refers to different iteration number. The thin vertical lines mark the median of the distributions.}\label{medians}
\end{figure*}

Because the signal-to-noise fluctuates randomly from one iteration to the next and the range in signal-to-noise values is large, we have chosen to consider the distribution throughout the entire model. The median value of this distribution tells us about how well converged the model is in general. We also consider the minimum value for the signal-to-noise for all levels and grid points. Figure \ref{medians} shows the signal-to-noise distributions for the energy levels 0 -- 6 individually and for all levels with a fractional population higher than $10^{-12}$. We use this cut-off, because levels which are less populated only add unwanted noise. The differently colored histograms show the distributions with increasing iteration number. On top of the distributions, the corresponding median values are marked. All medians are seen to increase with increasing iterations with the median of the least converged level 2 ending up at a signal-to-noise of 200 already at iteration 16. Still, the lowest signal-to-noise value of the entire model is as low as 20, which means that for that particular level we make an error of at most 5\%. We find this acceptable and therefore stop the calculation at this point. However, for better confidence, more photons and iterations can be used at the cost of calculation time. 

Another test which the user can perform in order to evaluate the convergence, is to plot the spatial distribution of the 10\% least converged points (or, for instance, all points with a signal-to-noise less than 100) and make a statistical comparison with the global point distribution to see if the least converged points in any way are associated with particular regions of the model or if the least converged points simply are a random subset to the entire grid.

One particular situation that the user should be aware of is the very highly opaque regime. In this regime the radiation transfer problem becomes similar to the diffusion problem and this can cause a very slow drift of the populations toward the correct solution. This drift may be so slow that the populations seem converged and this problem is inherent to radiation transfer codes.

\subsection{Ray-tracing}
After convergence has been reached, the code ray-traces lines of sight through the model in order to obtain an image cube of the radiation that escapes from the surface. The user provides the information on the source distance, source velocity, source orientation, and image resolution and units. The orientation parameters are slightly more complicated than for 2D codes, where an inclination and position angle are enough to set the source orientation. In \emph{LIME} we also have a source rotation which allows us to view a 3D model from any direction, using the matrix
\begin{eqnarray}
R_{\theta,\phi} = 
\left( \begin{array}{ccc}
cos(\phi) & 0 & -sin(\phi) \\
sin(\theta)sin(\phi) & cos(\theta) & sin(\theta)cos(\phi) \\
cos(\theta)sin(\phi) & -sin(\theta) & cos(\theta)cos(\phi) \end{array} \right),
\end{eqnarray}
where $\theta$ is the traditional inclination (0: face-on, $\pi/2$: edge-on) and $\phi$ is the azimuthal rotation. Rotation in the image plane is done afterwards, simply by rotating the image cube.

For the raytracing we let the photons move in straight lines, rather than jumping from grid point to grid point. In this part of the code we therefore do not make use of the Delaunay triangulation but rather the Voronoi diagram. The entire volume of the Voronoi cell is represented by the populations of the corresponding grid point and so integration of Eq.~\ref{radtran} becomes a matter of stepping through the source model and figuring out in which Voronoi cell the photon is. From an algorithmic point of view, this comes down to a simple sorting, not of thousands of cells, but only of, on average, the 16 neighboring cells. The step size is chosen as a fraction of the cell size in order to avoid accidentally missing a cell by stepping over it. Also, we need to sample variations in the velocity field across the cell in order to get a smooth spectrum. This is a very fast process compared to moving through a regular grid, but not as fast as moving along the Delaunay lines, which is why this transport method is not used when determining the level populations. 

Apart from the images, which are output in standard FITS format, \emph{LIME} can output a number of model diagnostics that are also obtained during the ray-tracing. This includes the opacity and intensity per Voronoi cell as seen along the line of sight. This information can be used to plot the $\tau=3$ surface and identify the origin of the radiation for various transitions and tracers. It is also possible to dump the grid to a text file together with the populations and physical quantities so that the grid structure and excitation temperature as well as the input model itself can be explored visually.

\subsection{Benchmark: 1D collapse model}
A standard problem in radiation transfer benchmarking is the spherical Shu-collapse model~\citep{shu1977}. This model describes a gaseous envelope that undergoes a gravitational inside-out collapse to form a protostar. In this example we use the particular formulation of the problem found in the code comparison project by~\citet{vanZadelhoff2002} so that we may compare our result to those of the codes that participated in that project\footnote{http://www.strw.leidenuniv.nl/astrochem/radtrans}. The molecule in consideration is HCO$^+$. Eight different codes were tested against each other, both in terms of performance and their solutions to the problem. In the present test we will only compare the solutions because hardware and compilers have changed too much in the last ten years for a performance comparison to make sense.
\begin{figure}
\includegraphics[width=8.8cm]{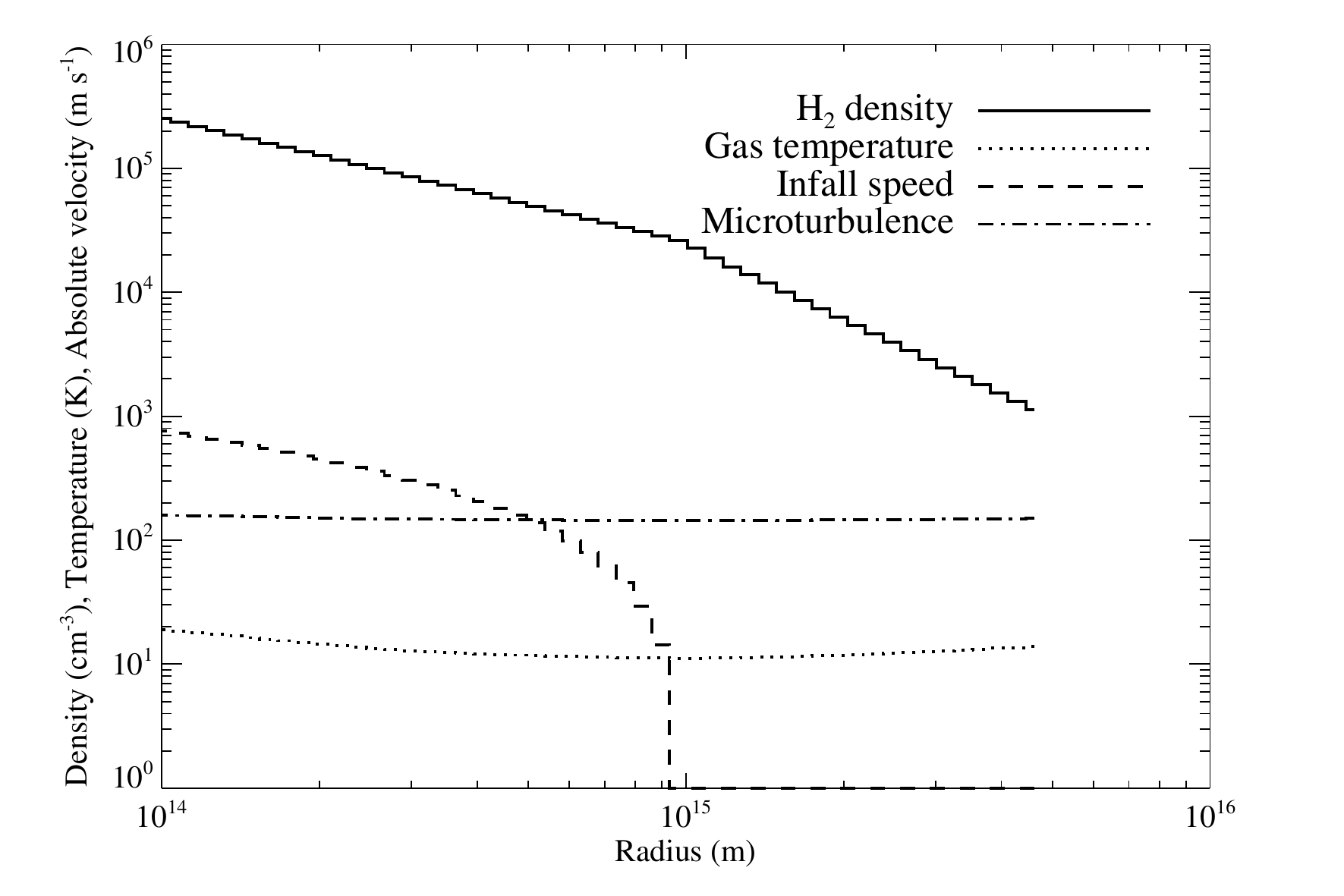}
\caption{The model setup for the 1D collapse benchmark test.}\label{zadelhoff_0}
\end{figure}

\begin{figure}
\includegraphics[width=8.8cm]{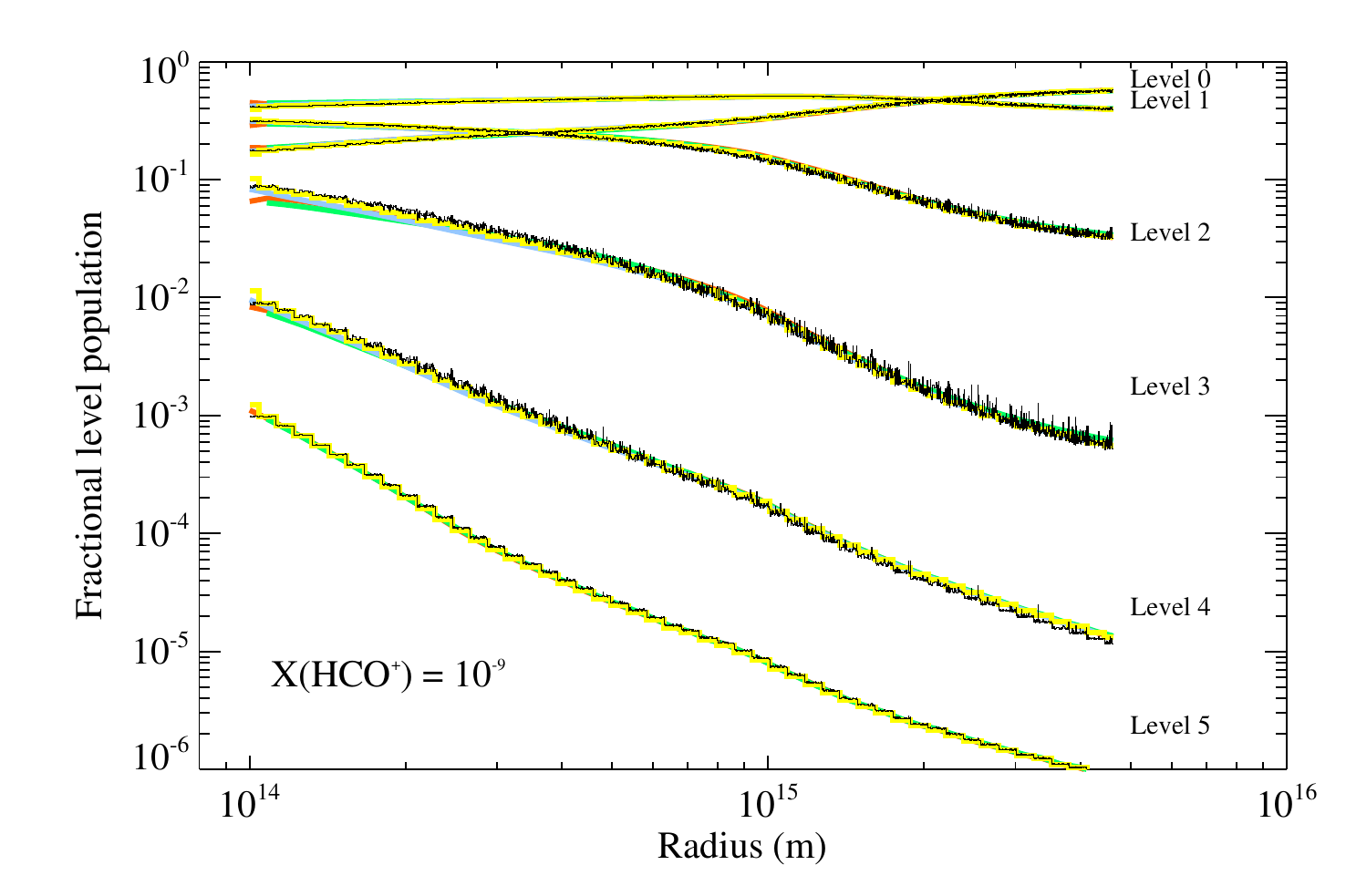}\\
\includegraphics[width=8.8cm]{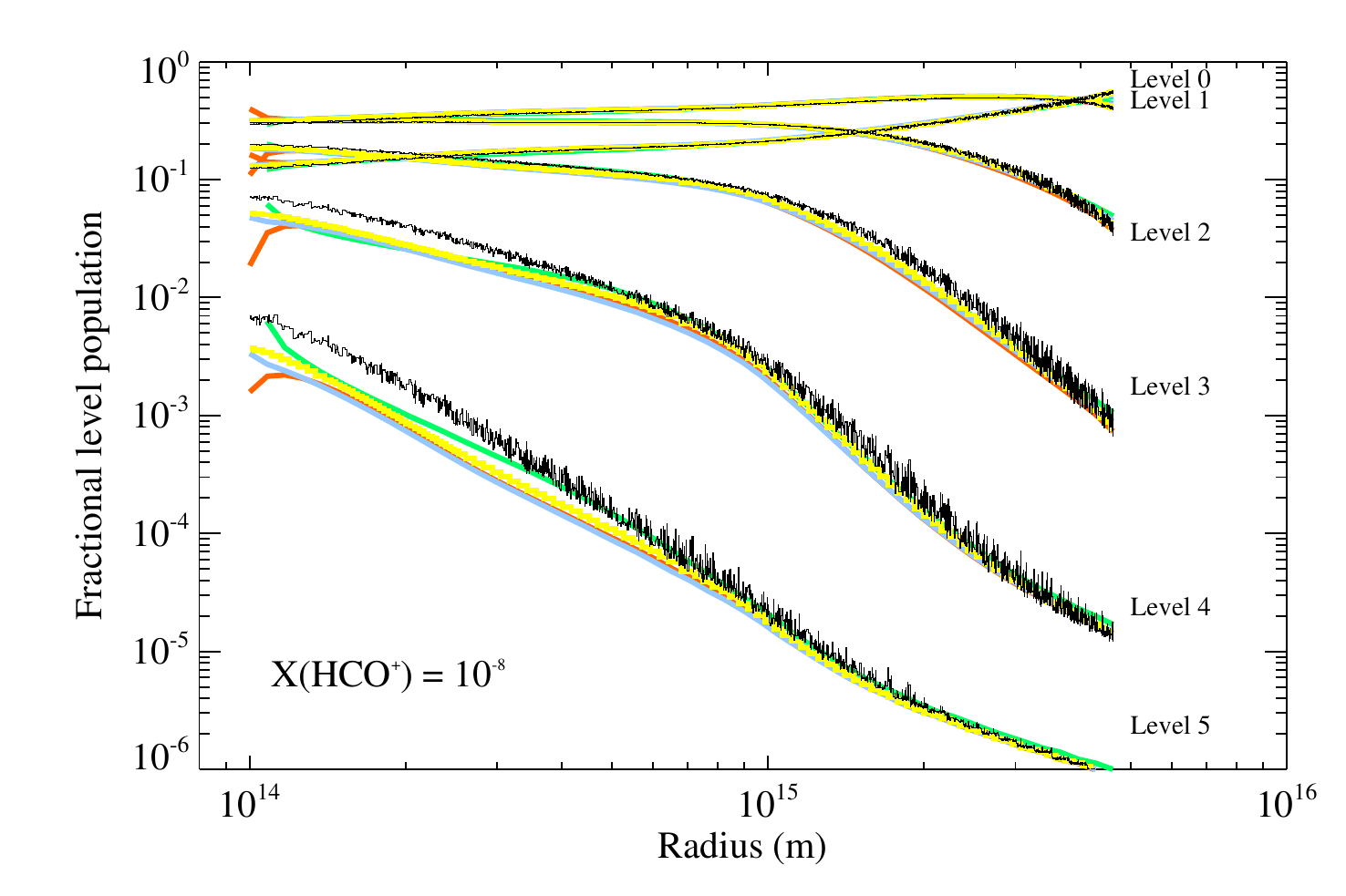}
\caption{Comparison of solutions to a 1D collapse model between different codes. The blue, red, and green curves are different solutions from the RADTRANS code comparison project. The yellow curve is the \emph{RATRAN} solution and the black line is the solution from \emph{LIME}.}\label{zadelhoff_1}
\end{figure}

The model is already provided as a logarithmically spaced table describing 50 grid cells, shown in Fig.~\ref{zadelhoff_0}. Since \emph{LIME} does not use regular grids, we interpolate this table and sample it randomly. For the comparison of solutions we have made a plot of the fractional level populations after convergence has been reached. There are two problems, denoted 2A and 2B in the paper by~\citet{vanZadelhoff2002}, corresponding to an optically thin and thick case. The fractional abundance of HCO$^+$ is 10$^{-9}$ and 10$^{-8}$ respectively. We show the result of both tests in Fig.~\ref{zadelhoff_1}. In this figure the fractional population of the ground state and the first four excited levels are shown. All eight codes participating in the RADTRANS comparison project agree very well. We have included only three of the eight solutions here (Doty, Dullemond, and Juvela) as well as the solution of \emph{RATRAN} in its present version. The black fluctuating line is the \emph{LIME} solution and is seen to agree very well with the established codes. In our 3D code, two points may be at the same distance from the center, but separated by a large angle and they may not be radiatively connected, especially not in high opacity models. These grid points may not see exactly the same radiation field because of the random nature of the grid and photon transport and therefore the level populations may not necessarily be exactly the same. There are no fluctuations in the results of the 1D codes because only a single cell with a single solution exist at a given radius. If we take the \emph{RATRAN} solution to be the expected solution we can calculate reduced $\chi^2$ values,
\begin{eqnarray}
\chi^2 = \frac{1}{N-1} \sum \frac{(n(r)_{LIME} - n(r)_{RATRAN})^2}{\sigma^2}
\end{eqnarray}
for each of the 6 levels of the \emph{LIME} solution which gives us $\chi^2_{l=0...5}\approx\{1.08, 1.11, 1.02, 1.17, 1.00, 1.00\}$ for the optically thin case and $\chi^2_{l=0...5}\approx\{1.06, 3.12, 1.03, 2.08, 1.28, 1.15\}$ for the optically thick case. The comparison is slightly worse for the optically thick case, but here we also see greater variation between the established codes.

\begin{figure*}
\begin{center}
\includegraphics[width=8.2cm]{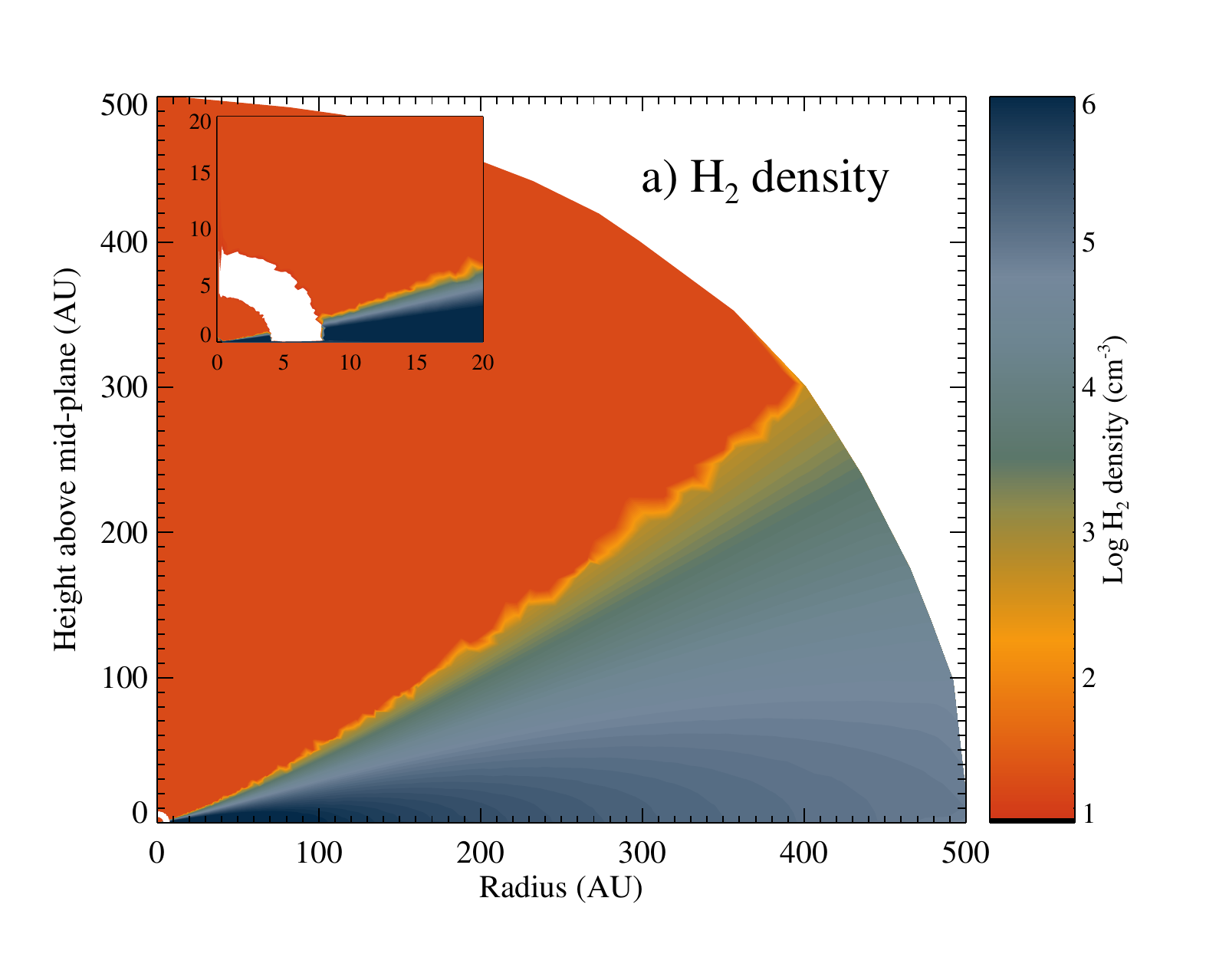}
\includegraphics[width=8.2cm]{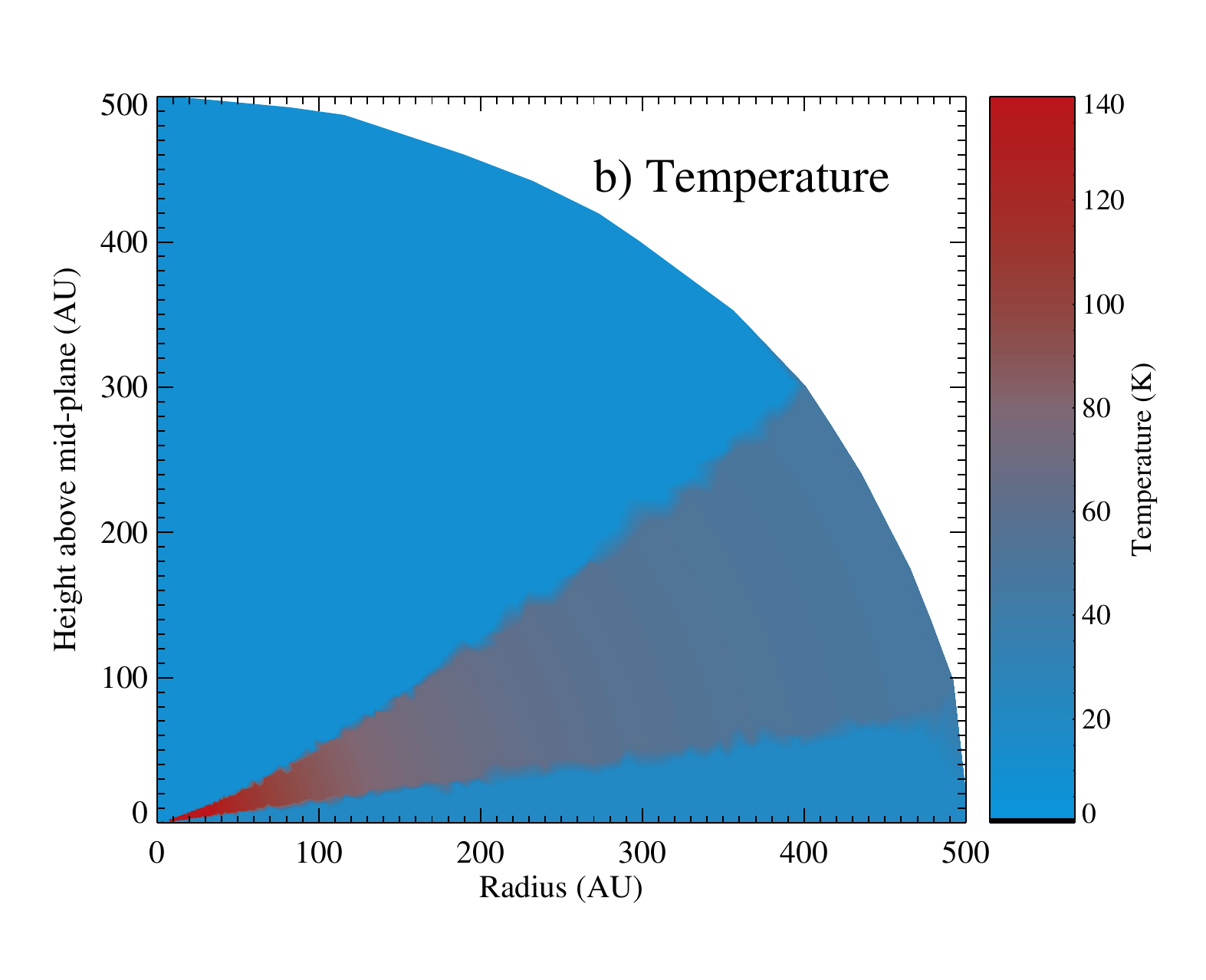}\\
\includegraphics[width=8.2cm]{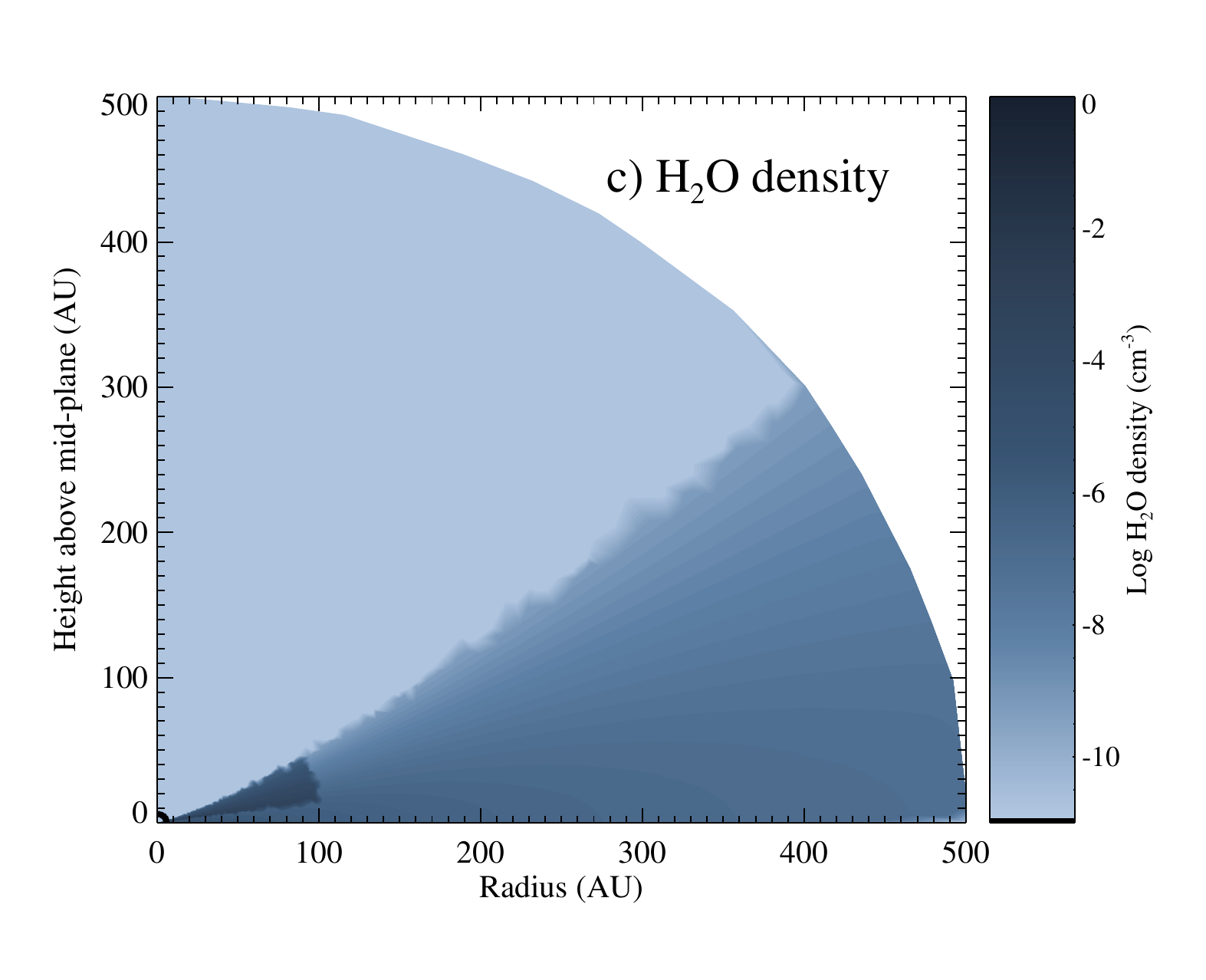}
\includegraphics[width=8.2cm]{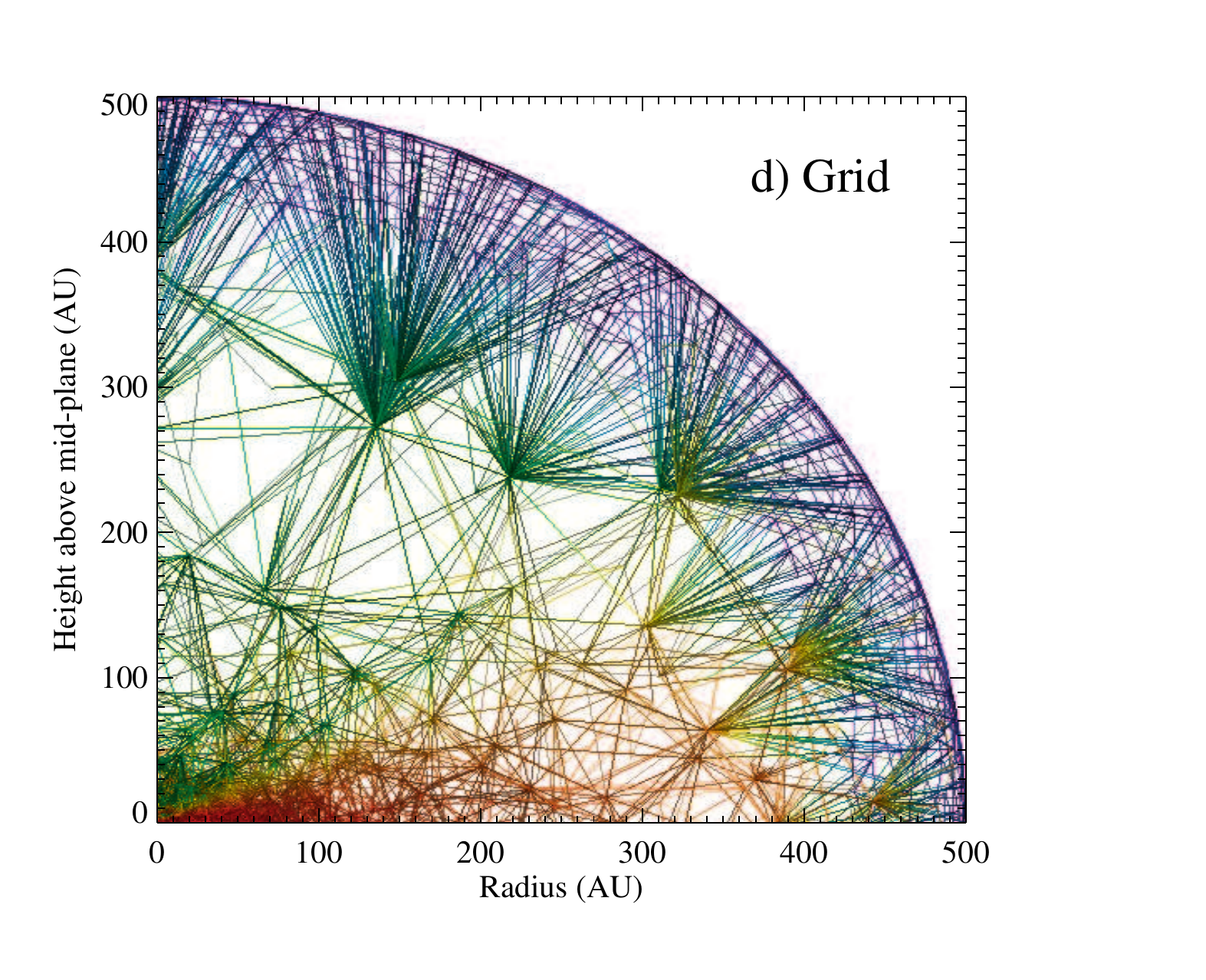}
\end{center}
\caption{Visualization of the upper right quadrant of the disk model described in Sect.~\ref{examples}. Panel a) shows the H$_2$ density with the insert showing the gap (shell) carved out around 5 AU. Panel b) shows the temperature and c) shows the molecular density which has a discontinuity around 90 K where water freezes out. The last panel, d), shows a cut through the grid, color coded according to the density.}\label{diskmod}
\end{figure*}

\section{Example}\label{examples}

The example we present is a typical 2D hydrostatic protoplanetary disk model with a cold and dense mid-plane. Such models are numerously found in the literature \citep[e.g.,][]{chiang1997, dullemond2004, robitaille2006} but here we use a simple analytic toy model for illustrative purposes. The density structure is given by
\begin{eqnarray}
n(r,z) = n_0 (r/r_0)^{-1.5} e^{-(z/h)^2},
\end{eqnarray}
where
\begin{eqnarray}
h= \sqrt{\frac{2 T\ k_B\ r^3}{G M_*}}.
\end{eqnarray}
We consider HCO$^+$, H$_2$O, and CH$_3$OH gas at a fractional abundance of 2$\times 10^{-9}$ with respect to the H$_2$ density. With these three species we illustrate the possibility of utilizing a large dynamic range in scales (HCO$^+$), very opaque models (H$_2$O), and multiple overlapping lines (CH$_3$OH). 

The temperature is given by a power-law
\begin{eqnarray}
T(r)=T_0 (r/r_0)^{-0.5}.
\end{eqnarray}
In a more realistic model the temperature would be calculated self-consistently based on the radiation properties of the central source and the temperature would drop toward the mid-plane because this region would be shielded from stellar radiation by the upper layers of the disk. In our example we mimic this effect by lowering the temperature in a wedge shaped region around the mid-plane to 20 K. By letting water freeze out at temperatures below 90 K, we can simulate a complex abundance structure often used in protoplanetary disks \citep{jonkheid2007,woitke2009}. The disk extends to 500 AU and the values for $n_0$ and $T_0$ are $10^8$ cm$^{-3}$ and 90 K at the radius of 100 AU. In addition we have added a 2 AU wide gap around a radius of 5 AU from the center. The disk is in Keplerian rotation. Figure~\ref{diskmod} shows the density, temperature, and H$_2$O density of our disk. Of other parameters that describes the disk are the turbulent velocity dispersion set to 150 ms$^{-1}$, stellar mass of 1 M$_\odot$, and a gas-to-dust ratio of 100. We use thin mantled grains with 10$^7$ years of coagulation and the resulting disk mass is 0.02 M$_\odot$.

To break the azimuthal symmetry and make the model fully 3D, we have placed a protoplanetary condensation in the gap. The protoplanet has the same qualitative properties as the one described in \citet{narayanan2006}. The protoplanet has been modeled by placing a spherical distribution of grid points at the desired spot and giving the grid points an H$_2$ number density of $2\times 10^{15}$ cm$^{-3}$, which, given a radius of 1000 Jupiter radii, results in a mass of about 1.4 M$_J$. The temperature of the condensation is kept at 150 K. 

Using this setup we have first made an edge-on view of the grid, which can be seen in panel d) of Fig.~\ref{diskmod}. The grid points (and their connections) are color coded according to density, where blue is lower density and red is higher density. The number of grid points in this simulation is 10$^4$ for the disk and 5000 for the planet. The disk is clearly seen to stand out in red, whereas the gap or the planet at 5 AU cannot be seen. The grid is written out in the beginning of the simulation and can be explored in 3D using readily available open-source tools (e.g., \emph{paraview}\footnote{http://www.paraview.org}).

\begin{figure*}
\includegraphics[width=8.8cm]{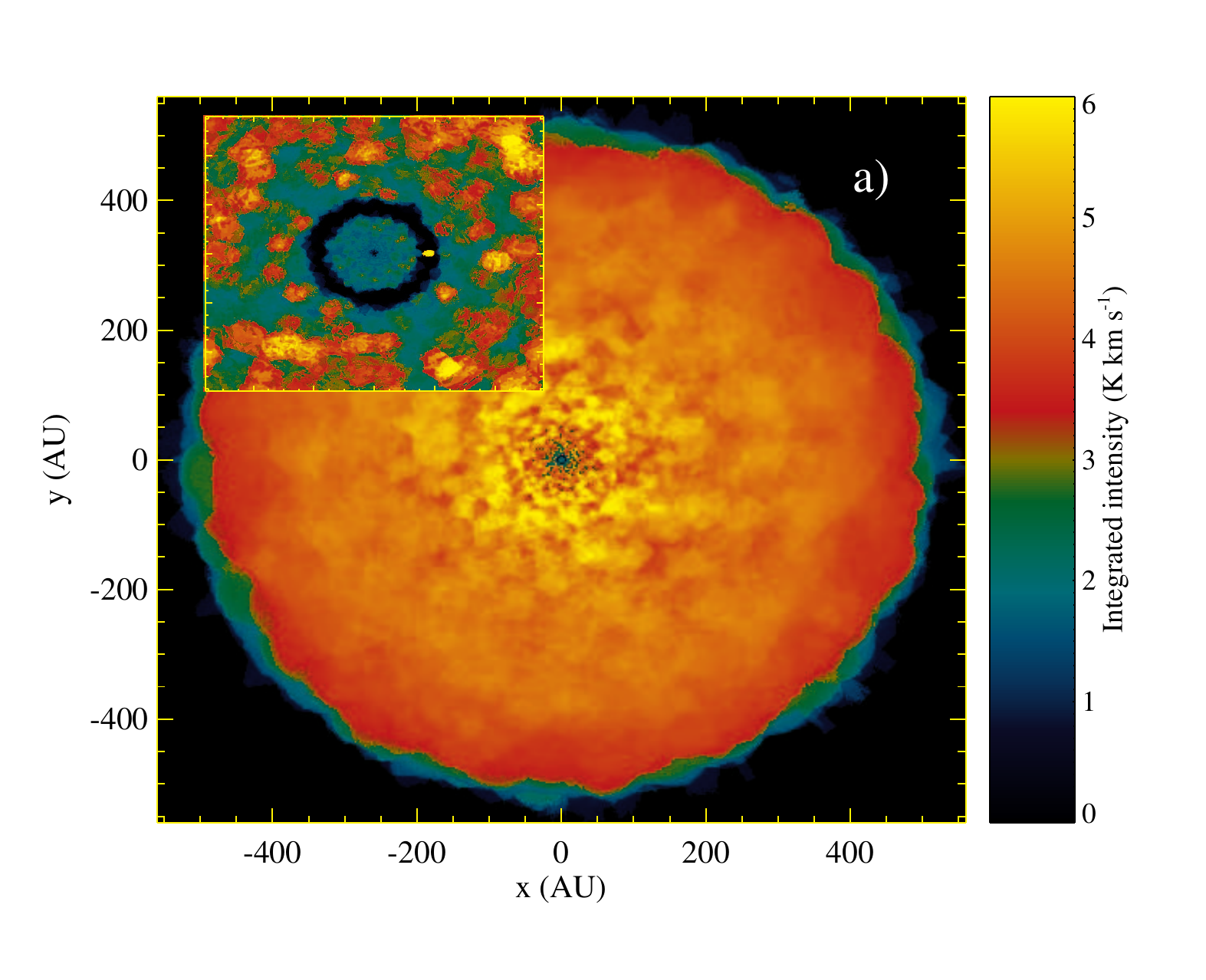} 
\includegraphics[width=8.8cm]{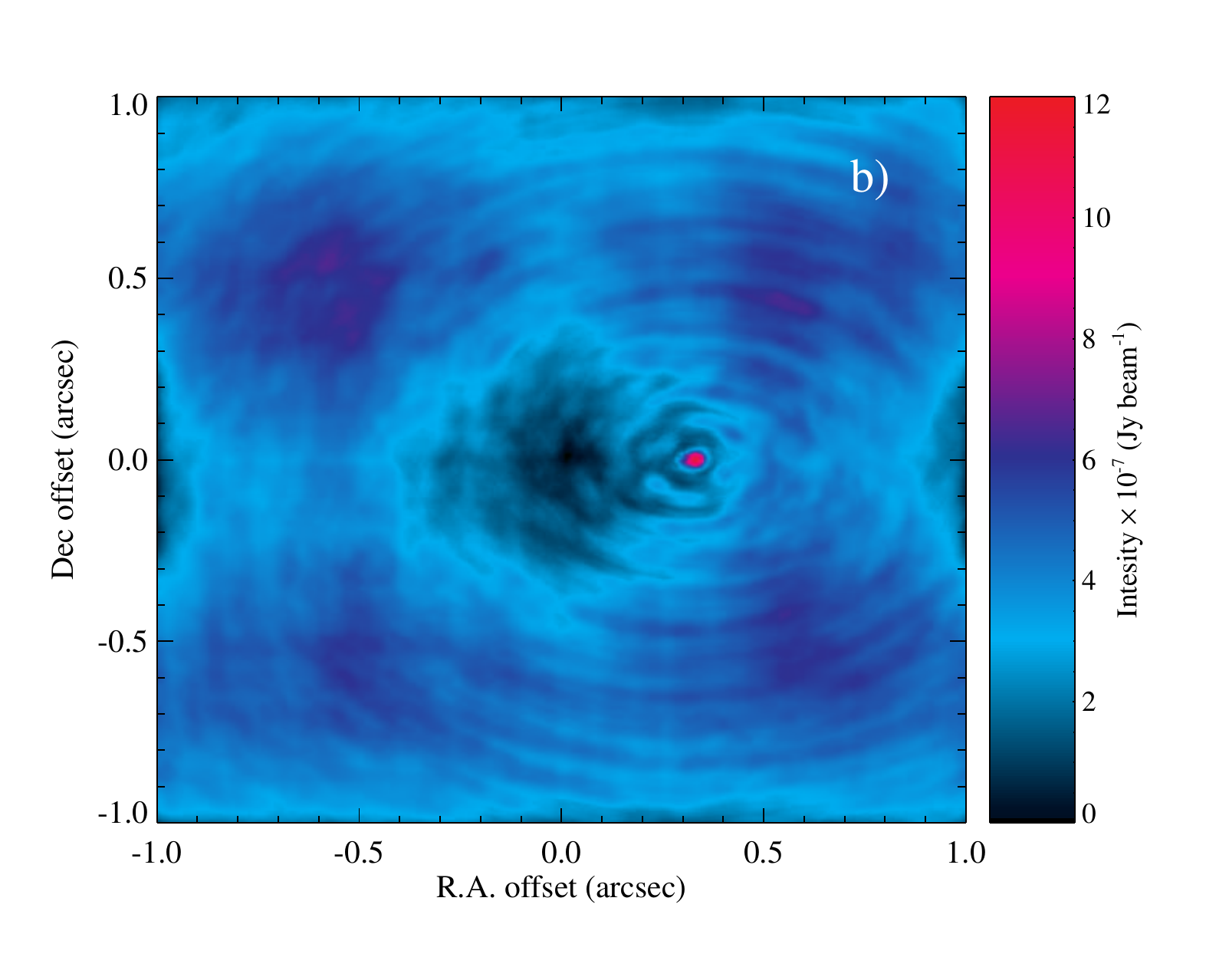}\\ 
\includegraphics[width=8.8cm]{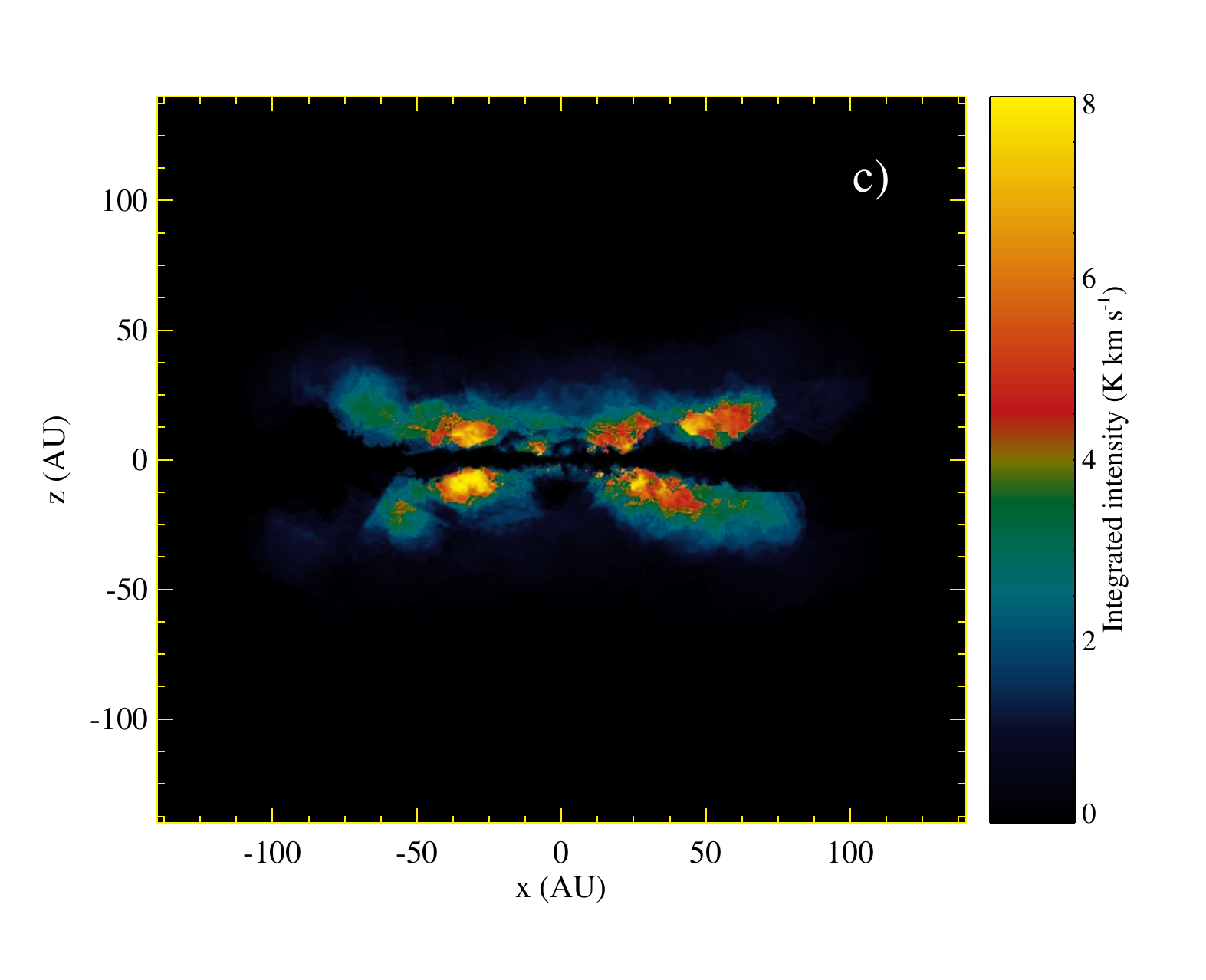} 
\includegraphics[width=8.8cm]{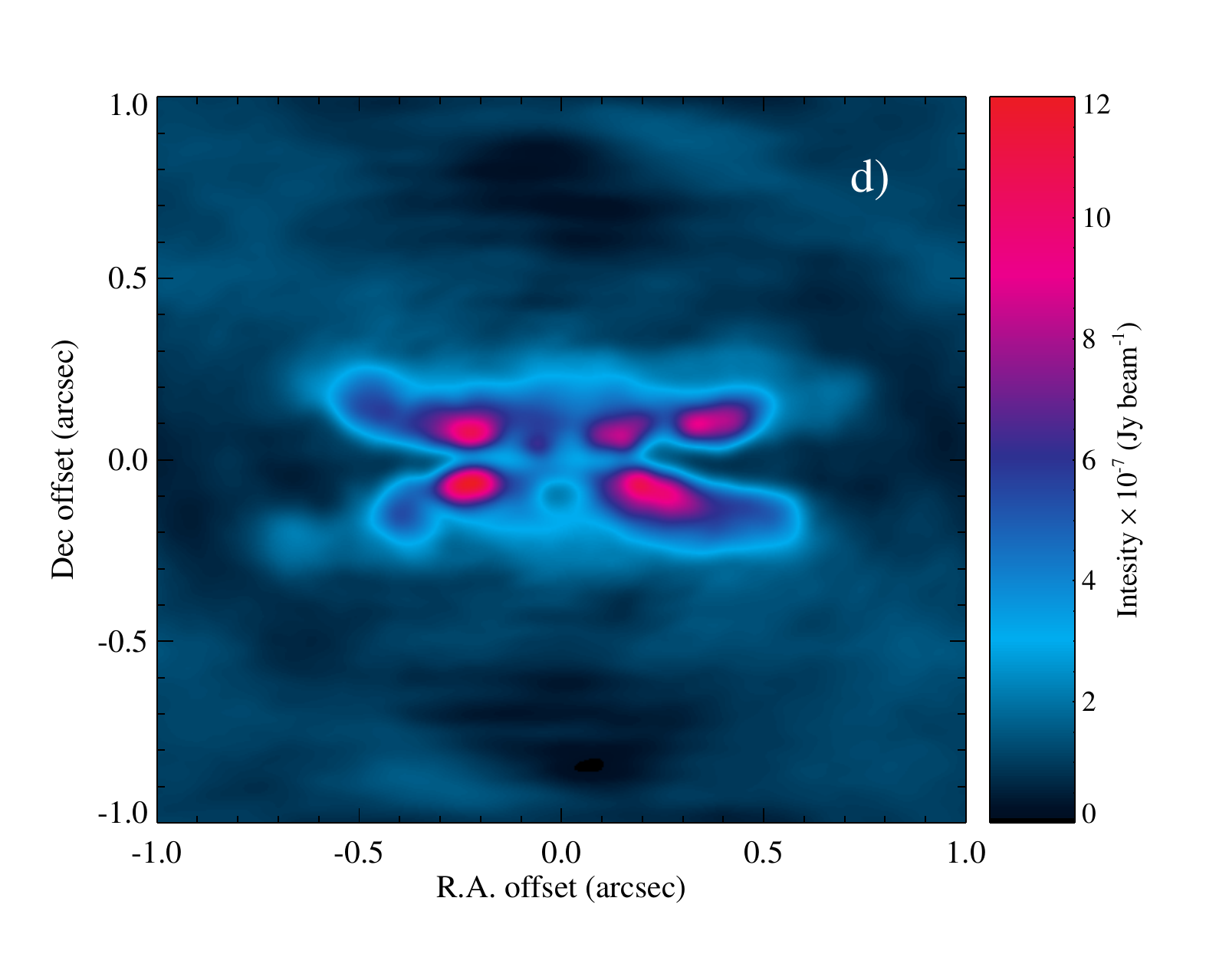}\\
\includegraphics[width=8.8cm]{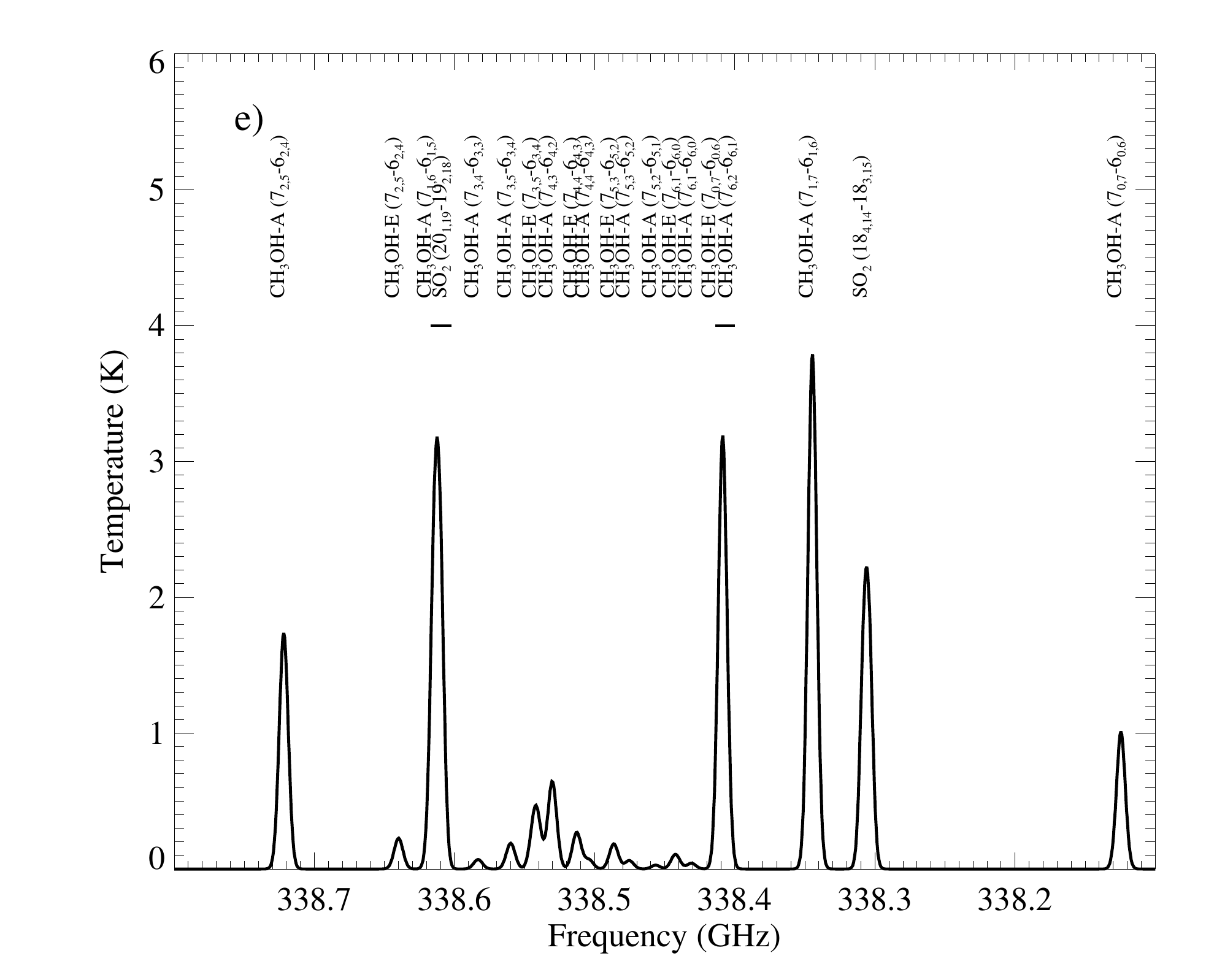} 
\includegraphics[width=8.8cm]{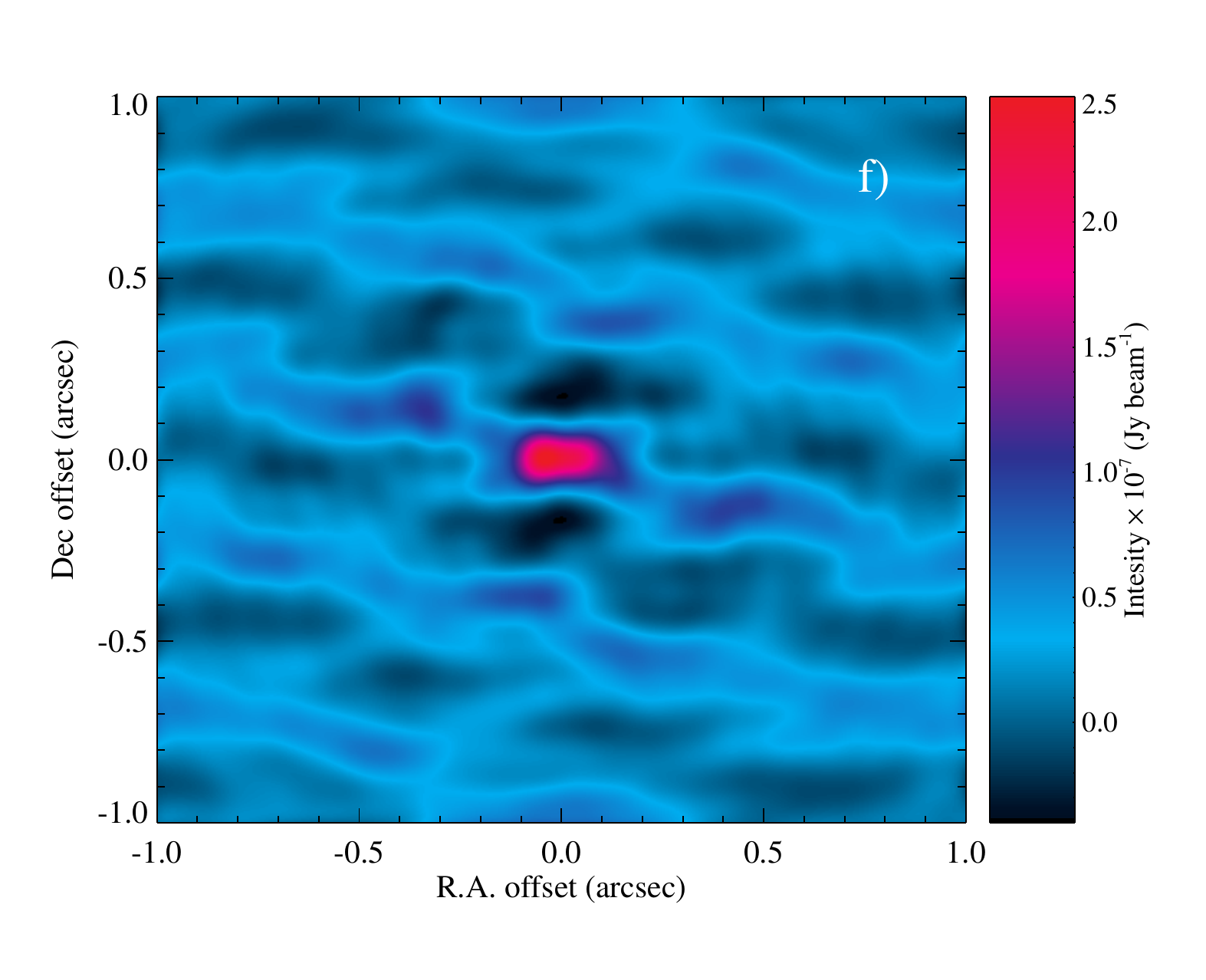}
\caption{The \emph{LIME} output (left column) and corresponding ALMA simulations (right column) of the disk model example presented in Sect.~\ref{examples}. Panel a) and b) show the disk face-on in HCO$^+$ $J=$7--6, panel c) and d) show the disk edge-on in H$_2$O $J=$ 3$_{13}$--2$_{20}$, and panel e) shows the average A- and E-CH$_3$OH spectrum as well as a couple of SO$_2$ lines which falls in this window. Panel f) is an ALMA simulation similar to the one in panel d), but for only 6 antennas. The ALMA simulations shown in panel b) and d) corresponds to 2 hour tracks with the full array. The noise level in both simulations is about $2.5 \times 10^{-7}$ Jy beam$^{-1}$. The simulation in panel f) is for a much more compact ALMA configuration with a noise level of $0.2 \times 10^{-7}$ Jy beam$^{-1}$.}\label{moment}
\end{figure*}

Panel a) of Fig~\ref{moment} shows the integrated HCO$^+$ $J=$ 7--6 emission as predicted by \emph{LIME}. This line has the frequency 624.2 GHz and falls in the ALMA band 9. The model is, in this case, viewed face-on and placed in the distance of 100 pc corresponding to the typical distance to the closest disks. The resolution of the image is 0.005", which is about an order of magnitude higher than the best ALMA resolution. In the insert, we have zoomed in on the inner most 10 percent of the disk, where the protoplanet can be seen as a small dot inside the ring-shaped gap. The intensity variations across the innermost parts of the disk are due to the grid structure and not a property of the model. Such fluctuations are unavoidable when using random grids, but can be remedied by smoothing the image or adding more grid points in the affected regions. Panel b) shows the same image, but using the ALMA simulator (i.e., the \emph{simData} task in the ALMA off-line data reduction package \emph{CASA}) to generate simulated (u,v)-spacings and realistic noise. The present simulation is for a two hour exposure in a very extended ALMA configuration. Given that the mean distance between grid points on the surface is less than 0.01 AU, this simulation spans almost 5 orders of magnitude in scales. 

In the panels c), d), and f) of Fig.~\ref{moment} we have flipped the disk so that we view it edge-on. In panel c) we show a model image for 183.3 GHz para-H$_2$O $J=$ 3$_{13}$--2$_{20}$ emission. This line falls in the ALMA band 5 and it is the only water transition available to ALMA. Only 6 band 5 receivers are planned so far, but here we show simulations of both a full array of band 5 receivers (panel d) and for the 6 planned receivers (panel f). In this edge-on view we see the lack of molecules in the mid-plane due to the simulated freeze-out. Both the \emph{LIME} model and the ALMA simulations has been continuum subtracted so that the water emission stands out. In the case of a full array of band 5 receivers (panel d), the water emission is resolved spatially. This is not the case when using only 6 antennas for the simulation (panel f) because the antennas needs to be placed in a much more compact configuration.

This is a very highly opaque model, with the optical depths going up to several thousands. This is handled quite well by \emph{LIME} although convergence takes 2-3 times more iterations than for HCO$^+$.  Because of the relative low abundance in this disk model, the water line does not maser in this particular example. Care should be taken when modeling potential maser lines with \emph{LIME} because it does not handle masering accurately. \emph{LIME} will produce a warning if the populations get inverted. 

The final example shows the multi-line ray-tracing capability of \emph{LIME}. Here we calculate the spectrum of both flavors of methanol (type A and E) and SO$_2$ around 338 GHz. We use the same disk model setup as above, but the lines are optically thin and we show the average spectrum over the disk so that result is relatively geometry independent. Figure~\ref{moment} panel e) shows the resulting spectrum in a 2 GHz windows centered around 338 GHz. In reality, methanol has many more lines in this window than what are shown here, but we are limited to the transitions for which the collision rates are know.

\section{Conclusions}\label{conclusion}
In this paper we have presented a new algorithm for solving the molecular excitation and radiation transfer problem in an arbitrary three dimensional geometry. The code uses a weighted stochastic point process to generate a random grid point distribution and its corresponding Delaunay triangulation on which the photon transport takes place. The code is well suited for calculating models with a large density contrast, which is particularly important for models of ALMA observations with a very high spatial resolution. Furthermore our code handles overlapping lines which is also going to be needed for proper modeling of ALMA data. Although emphasis has been put on making \emph{LIME} capable of modeling ALMA data, the code is also well suited for modeling data from other (sub-) millimeter interferometers (e.g., SMA and CARMA) as well as data from single-dish observatories (e.g., JCMT, APEX, Herschel Space Observatory).

In this paper we show a comparison between the \emph{LIME} code and a number of other molecular excitation and radiation transfer codes and we have applied \emph{LIME} to a model of a protoplanetary disk with a protoplanetary condensation.

\begin{acknowledgements}
The authors would like to acknowledge Ruud Visser for for thorough testing of the LIME code and for valuable feedback and suggestions. The authors would also like to acknowledge Kees Dullemond for helpful discussions on development of the LIME code. MRH acknowledges support from NWO grant 639.042.404. \end{acknowledgements}
\bibliographystyle{aa}
\bibliography{references}
\end{document}